\documentclass[proceedings]{JHEP} 

\usepackage{epsfig}			

\newbox\mybox
\newcommand\fverb{\setbox\mybox=\hbox\bgroup\verb}
\newcommand\fverbdo{\egroup\medskip\noindent\fbox{\unhbox\mybox}\ }
\newcommand\fverbit{\egroup\item[\fbox{\unhbox\mybox}]}

\font\beeg=cmr17 scaled 1600		
\newcommand\init[1]{\setbox\mybox=\hbox{{\beeg #1}~}%
		   \noindent\global\hangindent=\wd\mybox\global\hangafter-2%
		   \sc\smash{\llap {\lower 13.2pt \box\mybox}}}
\newcommand{\beq}{\begin{equation}}
\newcommand{\eeq}{\end{equation}}
\newcommand{\beqn}{\begin{eqnarray}}
\newcommand{\eeqn}{\end{eqnarray}}
\newcommand{\beqns}{\begin{eqnarray*}}
\newcommand{\eeqns}{\end{eqnarray*}}

\newcommand{\hm}{\hspace{-0.05cm}}

\newcommand{\intl}{\int\limits}

\newcommand{\e}{\epsilon}
\def\NP{{\it Nucl. Phys.}}
\def\PL{{\it Phys. Lett.}}
\def\PR{{\it Phys. Rev.}}

\def\PRL{{\it Phys. Rev. Lett.}}

\def\ZP{{\it Z. Phys.}}
\def\EPJ{{\it Europ. Phys. J.}}

\def\ea{{\it et al.}}
\def\Cl{Collaboration}
\def\pc{$\%$}

\def\sf{spectral function}
\def\sfs{spectral functions}

\def\as{$\alpha_s$}
\def\asm{$\alpha_s(m_\tau^2)$}

\def\ms{$m_s$}

\def\mss{$m_s(s)$}

\def\ee{$e^+e^-$}

\def\aqed{$\alpha(s)$}
\def\aqedZ{$\alpha(M_{\rm Z}^2)$}
\def\daqed{$\Delta\alpha(s)$}

\def\daqedhZ{$\Delta\alpha_{\rm had}(M_{\rm Z}^2)$}

\def\amuhad{$a_\mu^{\rm had}$}

\def\br{branching ratio}

\def\FOPTCI{$\rm FOPT_{\rm CI}$}
\def\Rt{$R_\tau$}

\def\RtS{$R_{\tau,S}$}

\def\RtVA{$R_{\tau,V/A}$}

\def\RtVpA{$R_{\tau,V+A}$}

\def\Rts{$R_\tau(s_0)$}

\def\RtVpAs{$R_{\tau,V+A}(s_0)$}

\def\ie{{\it i.e.}} 
 
\def\via{via} 

\title{Hadronic $\tau$ decays and QCD}

\author{Michel Davier\\
	Laboratoire de l'Acc\'el\'erateur Lin\'eaire\\
        IN2P3/CNRS et Universit\'e de Paris-Sud\\
        91898 Orsay, France\\
	E-mail: \email{davier@lal.in2p3.fr}}

\abstract{Hadronic decays of the $\tau$ lepton provide a clean source to
study hadron dynamics in an energy regime dominated by resonances, with the
interesting information captured in the spectral functions. Recent results 
on exclusive channels are reviewed. Inclusive spectral functions are the
basis for QCD analyses, delivering an accurate determination of the strong
coupling constant and quantitative information on nonperturbative
contributions. Strange decays yield a determination of the strange quark
mass.}

\begin{document} 

\maketitle 


\section{Introduction}

Hadrons produced in $\tau$ decays are borne out of the charged weak
current, {\it i.e.} out of the QCD vacuum. This property garantees
that hadronic physics factorizes in these processes which are then
completely characterized for each decay channel
by spectral functions as far as the total
rate is concerned . Furthermore, the produced
hadronic systems have $I=1$ and spin-parity $J^P=0^+,1^-$ (V) and
$J^P=0^-,1^+$ (A). The spectral functions are directly related to the
invariant mass spectra of the hadronic final states, normalized
to their respective branching ratios and corrected for the $\tau$
decay kinematics. For a given spin-1 vector decay, one has
\begin{eqnarray}
\label{eq_sf}
   v_1(s) 
   &\equiv&
           \frac{m_\tau^2}{6\,|V_{ud}|^2\,S_{\mathrm{EW}}}\,
              \frac{B(\tau^-\rightarrow {V^-}\,\nu_\tau)}
                   {B(\tau^-\rightarrow e^-\,\bar{\nu}_e\nu_\tau)} \nonumber \\
   & & \hspace{-1.2cm}        
              \times\frac{d N_{V}}{N_{V}\,ds}\,
              \left[ \left(1-\frac{s}{m_\tau^2}\right)^{\!\!2}\,
                     \left(1+\frac{2s}{M_\tau^2}\right)
              \right]^{-1}\hspace{-0.3cm}
\end{eqnarray}
where $V_{ud}$ denotes the CKM 
weak mixing matrix element and $S_{\mathrm{EW}}=1.0194\pm0.0040$ 
accounts for electroweak radiative corrections~\cite{marciano}. 

Isospin symmetry (CVC) connects the $\tau$ and $e^+e^-$ annihilation
spectral functions, the latter being proportional to the R ratio. For
example,
\begin{equation}
\label{eq_cvc}
 \sigma_{e^+e^-\rightarrow X^0}^{I=1}(s) \:=\:
 \frac{4\pi\alpha^2}{s}\,v_{1,\,X^-}(s)
\end{equation}
Radiative corrections violate CVC, as contained in the $S_{\mathrm{EW}}$
factor which is dominated by short-distance effects and thus expected
to be essentially final-state independent. 

Hadronic $\tau$ decays are then a clean probe of hadron dynamics in an
interesting energy region dominated by resonances. However, perturbative
QCD can be seriously considered due to the relatively large $\tau$ mass.
Many hadronic modes have been measured and studied, while some earlier
discrepancies (before 1990) have been resolved with high-statistics and
low-systematics experiments. Samples of $\sim 4 \times 10^5$ measured decays
are available in each LEP experiment and CLEO. Conditions for low 
systematic uncertainties are particularly well met at LEP: measured
samples have small non-$\tau$ backgrounds ($<1\%$) and large selection
efficiency ($92\%$), for example in ALEPH.

Recent results in the field are discussed in this report.

\section{Specific final states}

\subsection{Vector states}

The decay $\tau \rightarrow \nu_\tau \pi^- \pi^0$ is now studied with large
statistics of $\sim 10^5$ events. Data from ALEPH have been 
published~\cite{aleph_v}. New results from CLEO are now 
available~\cite{cleo_2pi} with the mass spectrum 
given in figure~\ref{cleo_2pi} 
dominated by the $\rho$(770) resonance. Good agreement is observed
between the ALEPH and CLEO data and the lineshape fits show strong
evidence for the contribution of $\rho$(1400) through interference with
the dominant amplitude. Fits also include a $\rho$(1700) contribution,
taken from $e^+e^-$ data as the value of the $\tau$ mass does not allow
$\tau$ data alone to tie down the corresponding resonance parameters.
Thanks to the high precision of the data, fits are sensitive to the exact
form of the $\rho$ lineshape, with a preference given to the 
Gounaris-Sakurai parametrization~\cite{gounaris} over that of
K\"uhn-Santamaria~\cite{ksm}.

\smallskip
\FIGURE{\epsfig{file=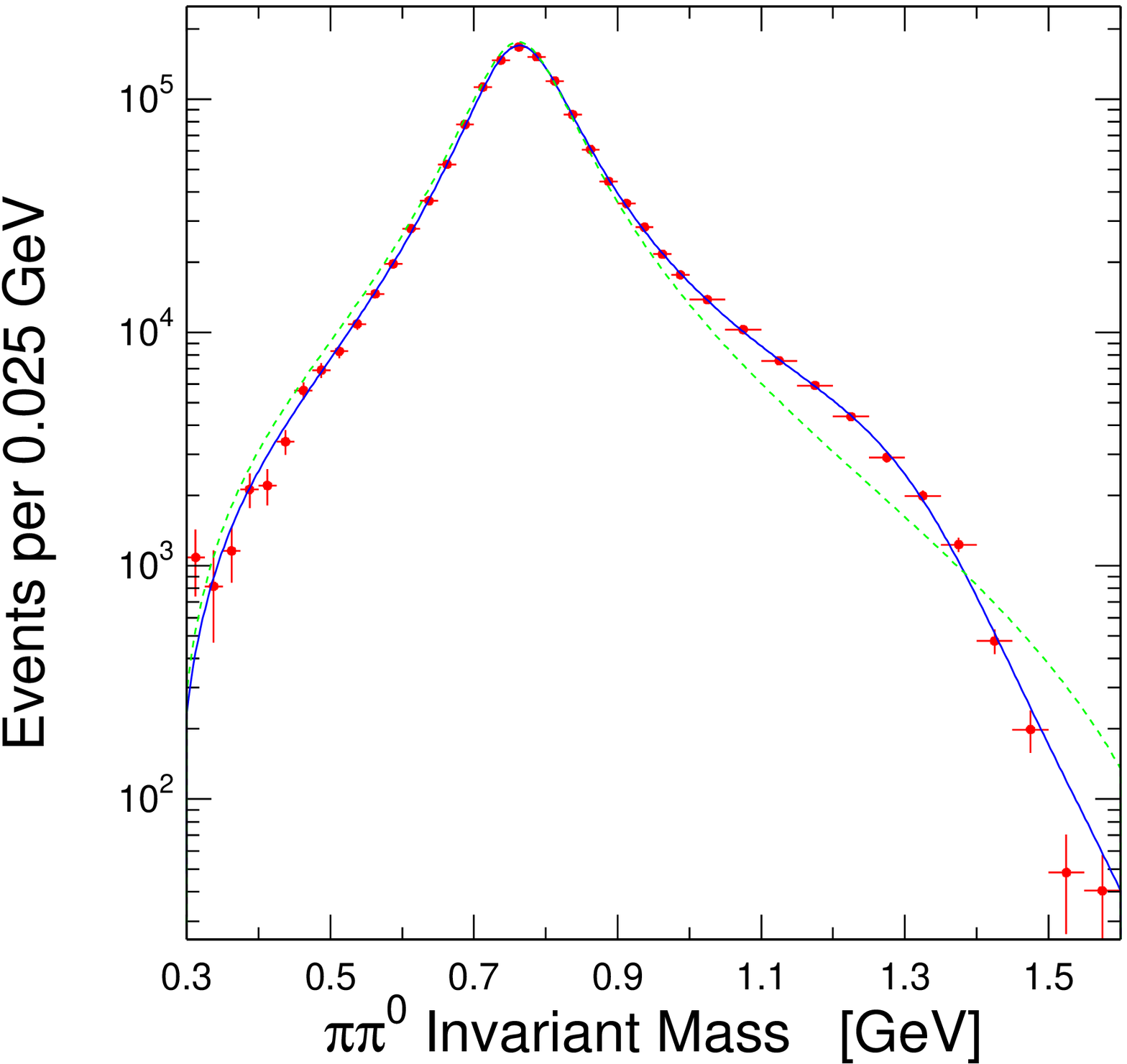,width=6cm}%
        \caption{Mass distribution in CLEO 
         $\tau \rightarrow \nu_\tau \pi^- \pi^0$ events. The solid curve
         overlaid is the result of the K\"uhn-Santamaria fit, while the 
         dashed curve has the $\rho(1400)$ contribution turned off.}%
	\label{cleo_2pi}}

The quality of data on $e^+e^- \rightarrow \pi^+ \pi^-$ has also recently
improved with the release of the CMD-2 results from Novosibirsk~\cite{cmd}.
A comparison of the mass spectrum as measured in $e^+e^-$ and $\tau$ data
is given in figure~\ref{aleph-ee_2pi} 
(for this exercise the $\rho-\omega$ interference
has to be artificially introduced in the $\tau$ data). Although the 
agreement looks impressive, it is possible to quantify it by computing 
a single number, integrating over the complete spectrum. It is convenient
for this to use the branching ratio 
$B(\tau \rightarrow \nu_\tau \pi^- \pi^0)$ as directly measured in $\tau$
decays and computed from the $e^+e^-$ spectral function under the
assumption of CVC. Using 
$B(\tau \rightarrow \nu_\tau h^- \pi^0)=(25.79 \pm 0.15)\%$~\cite
{heltsley} and subtracting out 
$B(\tau \rightarrow \nu_\tau K^- \pi^0)=(0.45 \pm 0.02)\%$~\cite{aleph_k},
one gets $B(\tau \rightarrow \nu_\tau \pi^- \pi^0)=(25.34 \pm 0.15)\%$,
somewhat larger than the CVC value using all available $e^+e^-$ data,
$B^{CVC}_{2 \pi}=(24.65 \pm 0.35)\%$~\cite{eidelman_HD}. This $1.8 \sigma$
discrepancy should be further investigated with a detailed examination of
the respective possible systematic effects, such as radiative corrections
in $e^+e^-$ data and $\pi^0$ reconstruction in $\tau$ data. CVC violations
are of course expected at some point: hadronic violation should be very
small ($\sim (m_u-m_d)^2/m_\tau^2$), while significant effects could arise
from long-distance radiative processes. Estimates show that the difference
between the charged and neutral $\rho$ widths should only be at the level 
of $(0.3 \pm 0.4)\%$~\cite{alemanyhd}.

\smallskip
\FIGURE{\epsfig{file=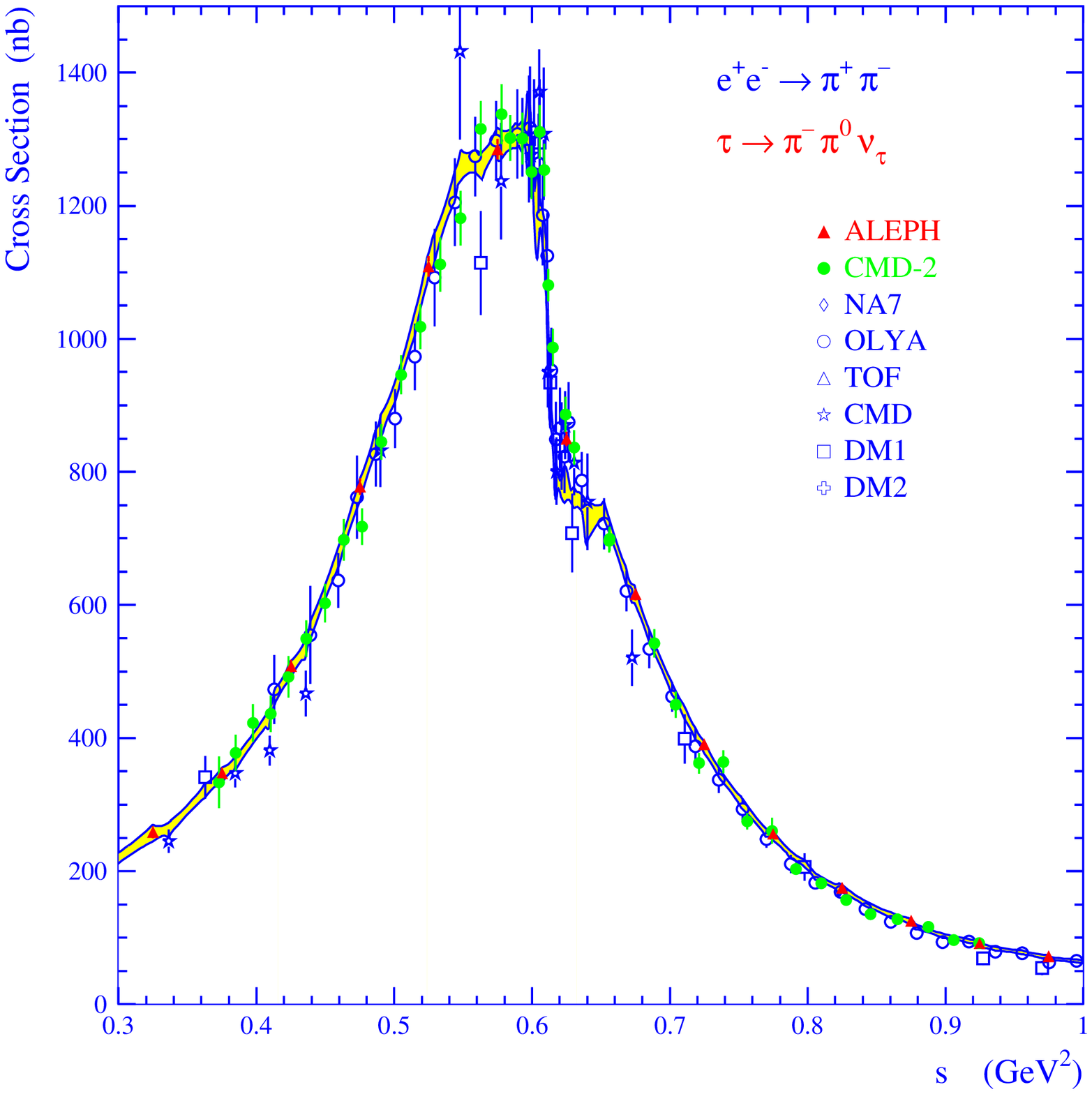,width=6cm,
         bbllx=70,bblly=150,bburx=560,bbury=710}%
        \caption{Cross section for $e^+e^- \rightarrow \pi^+ \pi^-$
          compared to the ALEPH $\tau$ data using CVC with $\rho-\omega$
          interference built in (shaded band).}%
	\label{aleph-ee_2pi}}

The $4 \pi$ final states have also been studied~\cite{aleph_v,cleo_4pi}.
Tests of CVC are severely hampered by large deviations between different
$e^+e^-$ experiments which disagree well beyond their quoted systematic
uncertainties. A new CLEO analysis studies the resonant structure in the
$3 \pi \pi^0$ channel which is shown to be dominated by $\omega \pi$ and
$a_1 \pi$ contributions. The $\omega \pi$ spectral function shown in
figure~\ref{cleo_4pi} 
is in good agreement with CMD-2 results and it is interpreted by
a sum of $\rho$-like amplitudes. The mass of the second state is however
found at $(1523 \pm 10)$MeV, in contrast with the value $(1406 \pm 14)$MeV
from the fit of the $2 \pi$ spectral function. This point has to be
clarified. Following a limit of $8.6\%$ obtained earlier by 
ALEPH~\cite{aleph_omega}, CLEO sets a new $95\%$ CL limit of $6.4\%$ for
the relative contribution of second-class currents in the decay
$\tau \rightarrow \nu_\tau \pi^- \omega$ from the hadronic angular decay
distribution.

\smallskip
\FIGURE{\epsfig{file=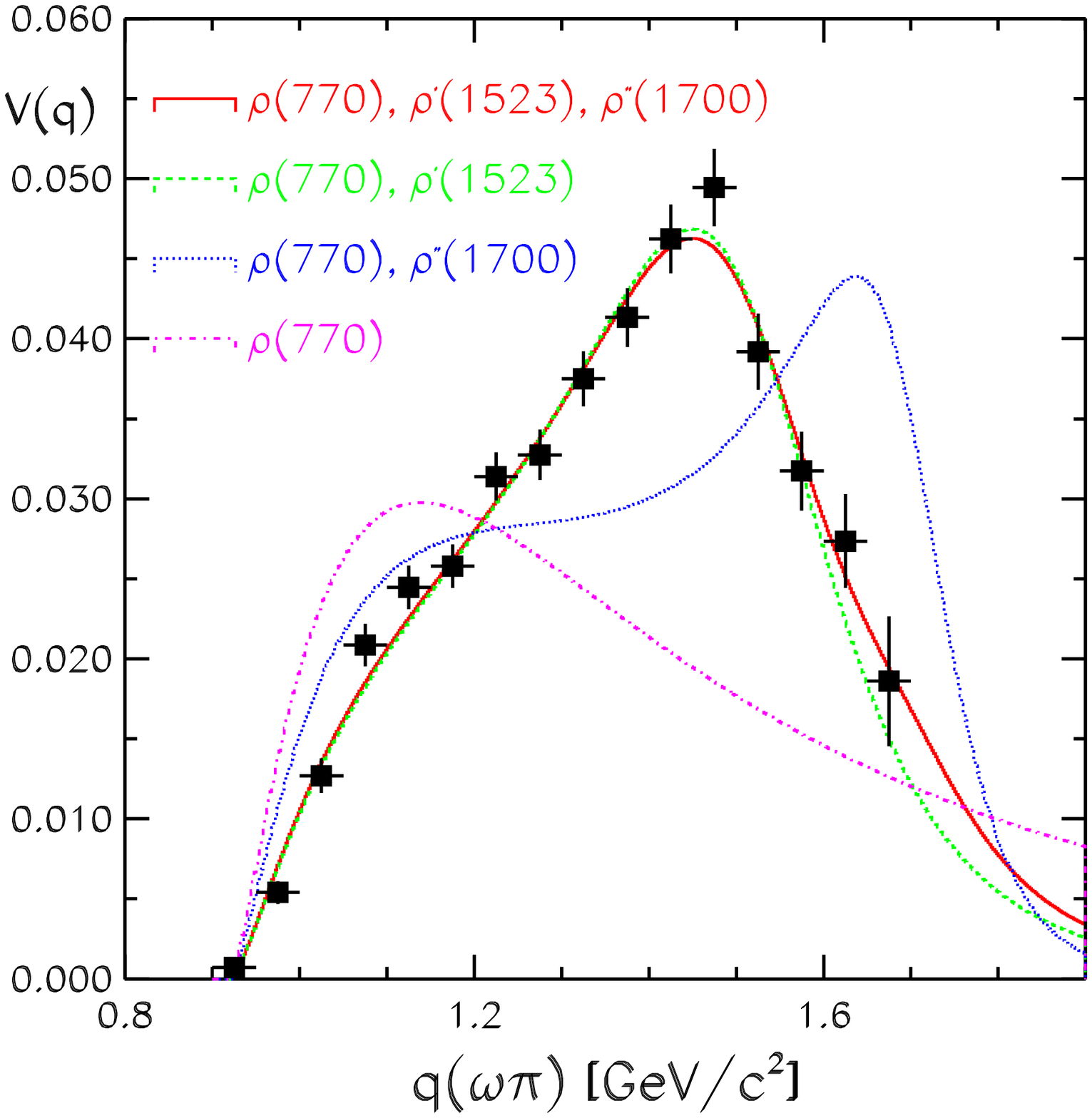,width=5cm}%
        \caption{Fits of the CLEO spectral function for
         $\tau \rightarrow \nu_\tau \pi^- \omega$.}%
	\label{cleo_4pi}}

\subsection{Axial-vector states}

The decay $\tau \rightarrow \nu_\tau 3\pi$ is the cleanest place to study
axial-vector resonance structure. The spectrum is dominated by the $1^+$ 
$a_1$ state, known to decay essentially through $\rho \pi$. A comprehensive
analysis of the $\pi^- 2\pi^0$ channel has been presented by CLEO. 
First, a model-independent determination of the hadronic structure 
functions gave no evidence for non-axial-vector contributions 
($<17\%$ at $90\%$ CL)~\cite{cleo_3pisf}. Second, a partial-wave
amplitude analysis was performed~\cite{cleo_3pi}: while the dominant 
$\rho \pi$ mode was of course confirmed, it came as a surprize that an 
important contribution ($\sim 20\%$) from scalars 
($\sigma$, $f_0(1470)$, $f_2(1270)$) was found in the $2\pi$ system.

The $a_1 \rightarrow \pi^- 2\pi^0$ lineshape is displayed 
in figure~\ref{cleo_3pi} where the opening 
of the $K^*K$ decay mode in the total $a_1$ width is clearly
seen. The derived branching ratio, $B(a_1 \rightarrow K^*K)=(3.3 \pm 0.5)\%$
is in good agreement with ALEPH results on the $K \bar K \pi$ modes which
were indeed shown (with the help of $e^+e^-$ data and CVC) to be 
axial-vector ($a_1$) dominated with $B(a_1 \rightarrow K^*K)=(2.6 \pm 0.3)\%$
~\cite{aleph_k}. No conclusive evidence for a higher mass state ($a'_1$)
is found in this analysis.

\smallskip
\FIGURE{\epsfig{file=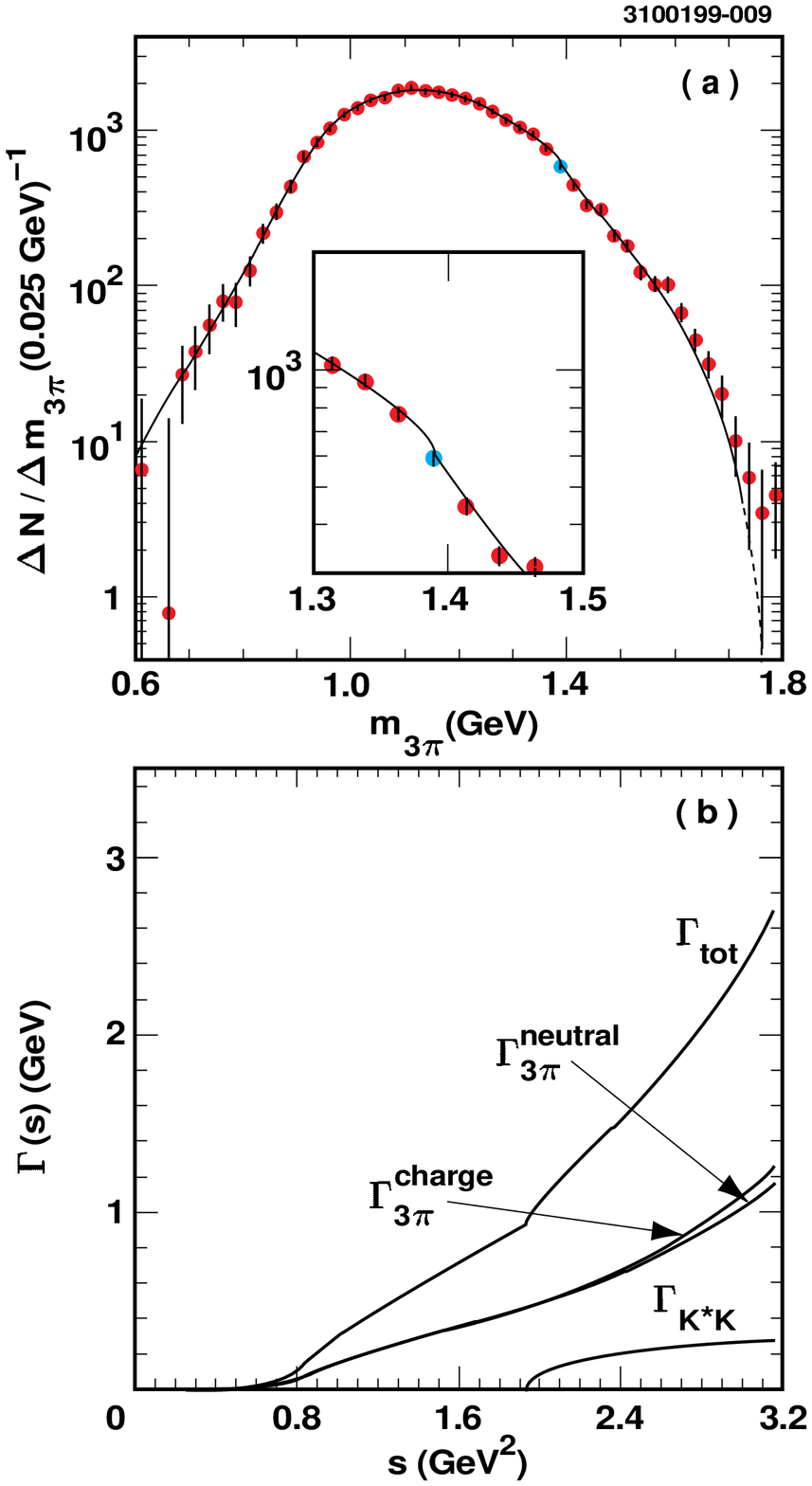,width=5cm} 
        \caption{(a) $3 \pi$ mass spectrum from CLEO in the $\pi^- 2\pi^0$ 
         channel with the lineshape from the resonance fit (zoom
         on the $K^*K$ threshold). In (b), the $\sqrt{s}$-dependent $a_1$  
         width is plotted with the different contributions considered.}%
	\label{cleo_3pi}}

\section{Inclusive spectral functions}

The $\tau$ nonstrange spectral functions have been measured by ALEPH
~\cite{aleph_v,aleph_a} and OPAL~\cite{opal}. The procedure requires a
careful separation of vector (V) and axial-vector (A) states involving
the reconstruction of multi-$\pi^0$ decays and the proper treatment of
final states with a $K \bar K$ pair. The $V$ and $A$ spectral functions are
given in figures~\ref{v_aleph_opal} and \ref{a_aleph_opal}, respectively.
They show a strong resonant behaviour, dominated by the lowest 
$\rho$ and $a_1$
states, with a tendancy to converge at large mass toward a value near
the parton model expectation. Yet, the vector part stays clearly above
while the axial-vector one lies below. Thus, the two spectral functions
are clearly not 'asymptotic' at the $\tau$ mass scale.

The $V+A$ spectral function, shown in figure~\ref{vpa_aleph} has a clear 
converging pattern toward a value above the parton level as expected in
QCD. In fact, it displays a textbook example of global duality, since
the resonance-dominated low-mass region shows an oscillatory behaviour around
the asymptotic QCD expectation, assumed to be valid in a local sense only
for large masses. This feature will be quantitatively discussed in the 
next section.
 
\smallskip
\FIGURE{\epsfig{file=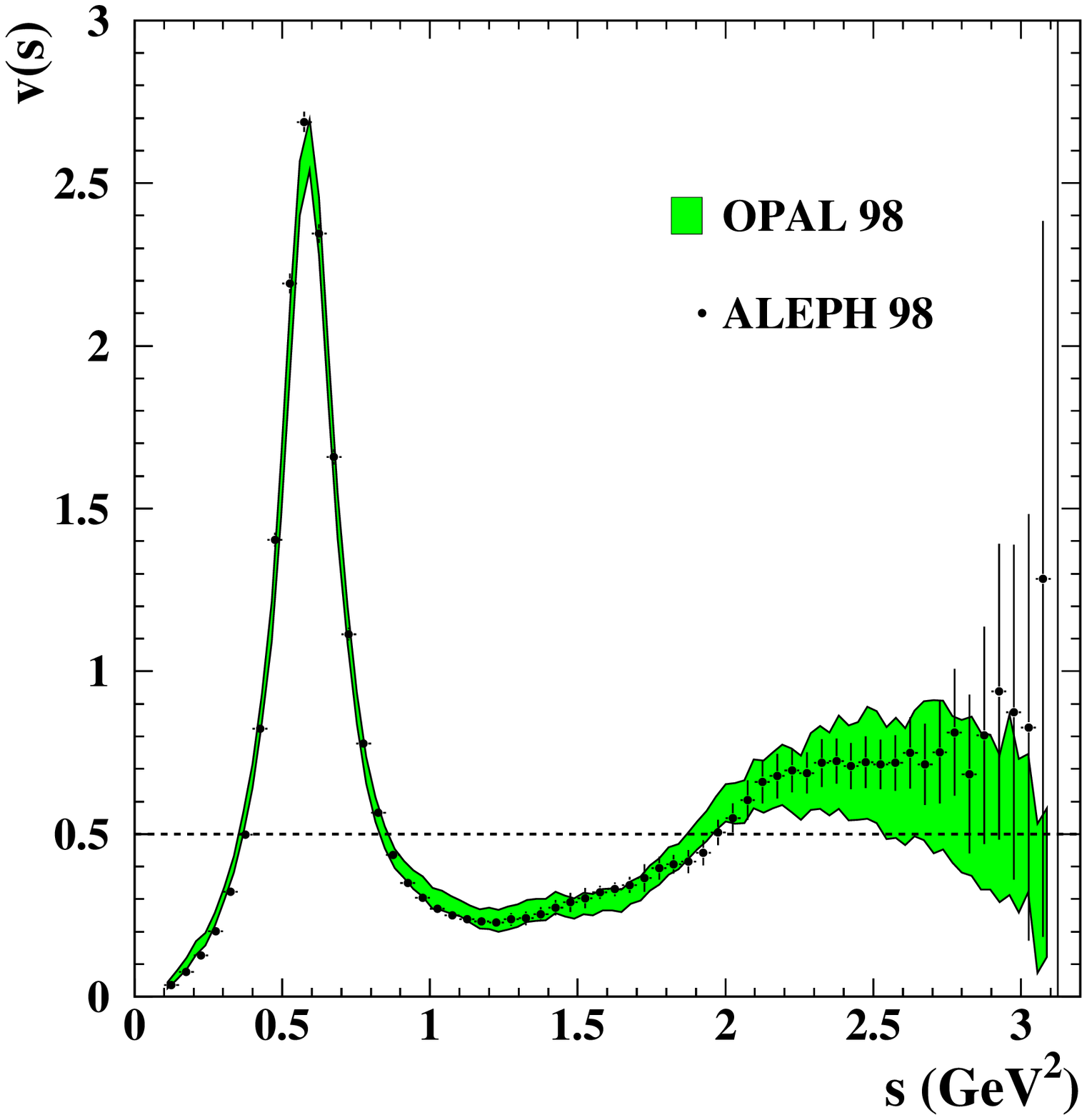,width=5cm,
         bbllx=70,bblly=150,bburx=560,bbury=650}%
        \caption{Inclusive nonstrange vector spectral function from 
         ALEPH and OPAL. The dashed line is the expectation from the naive 
         parton model}%
	\label{v_aleph_opal}}
\smallskip

\FIGURE{\epsfig{file=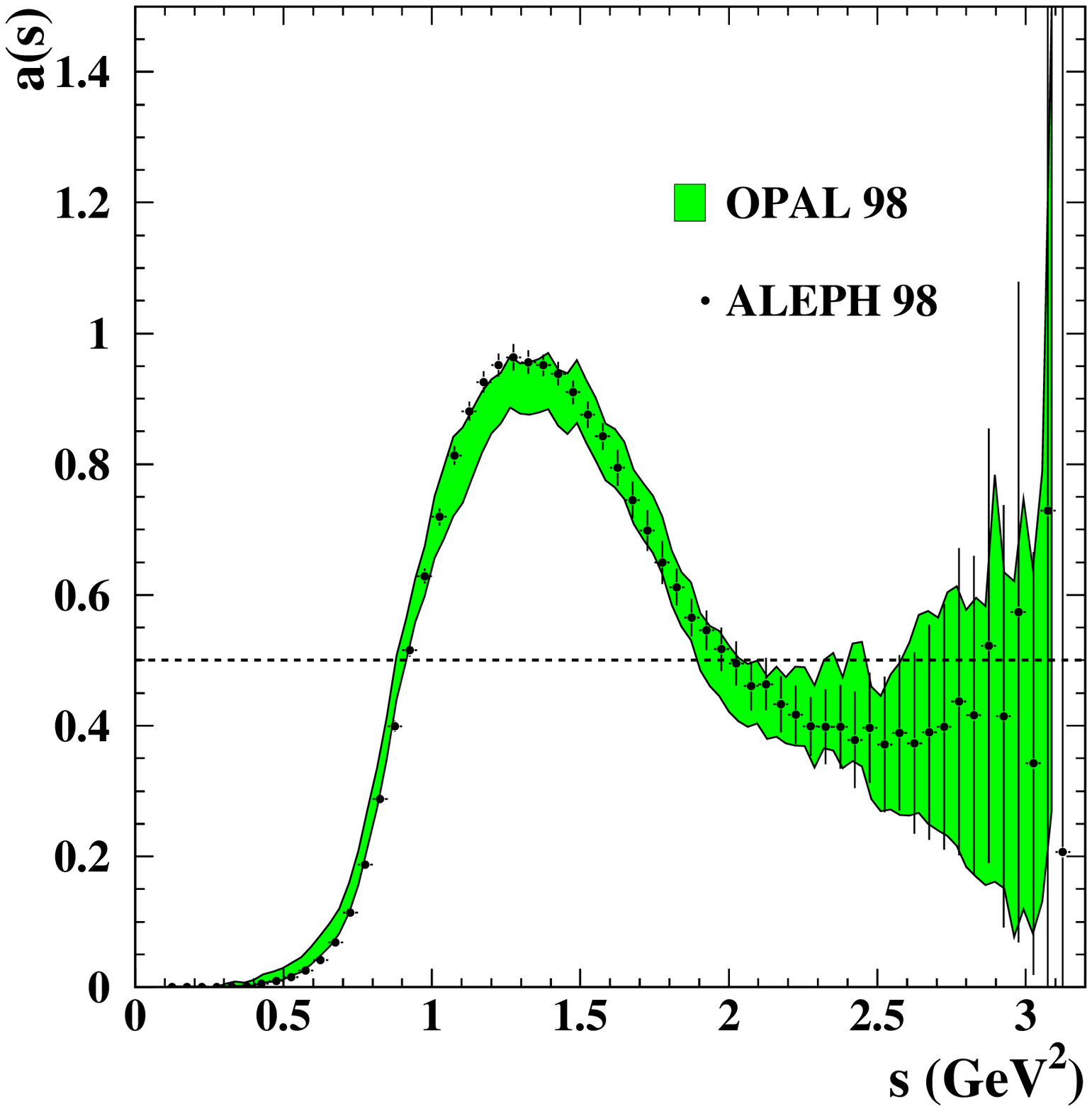,width=5cm,
         bbllx=70,bblly=150,bburx=560,bbury=650}%
        \caption{ Inclusive nonstrange axial-vector spectral function from 
         ALEPH and OPAL. The dashed line is the expectation from the naive 
         parton model.}%
	\label{a_aleph_opal}}
\smallskip

\FIGURE{\epsfig{file=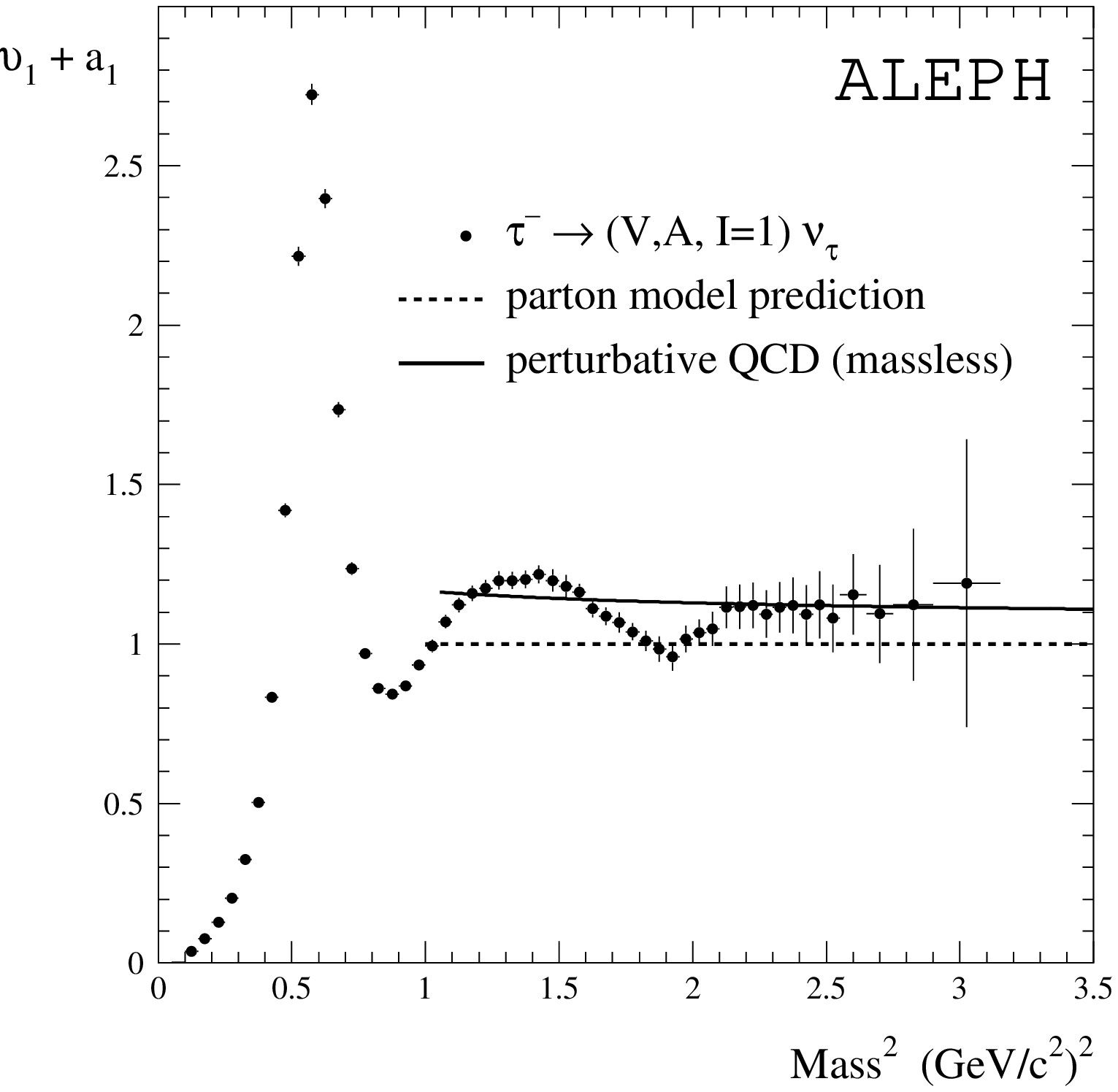,width=6cm}%
        \caption{ Inclusive $V+A$ nonstrange spectral function from ALEPH. 
         The dashed line is the expectation from the naive parton model,
         while the solid one is from massless perturbative QCD using
         $\alpha_s(M_{\rm Z}^2)=0.120$.}%
	\label{vpa_aleph}}

\section{QCD analysis of nonstrange $\tau$ decays}

\subsection{Motivation}

The total hadronic $\tau$ width, properly normalized to the known leptonic
width,

\beq
     R_\tau = \frac{\Gamma(\tau^-\rightarrow{\rm hadrons}^-\,\nu_\tau)}
                   {\Gamma(\tau^-\rightarrow e^-\,\bar{\nu}_e\nu_\tau)}
\eeq
should be well predicted by QCD as it is an inclusive observable. Compared
to the similar quantity defined in \ee\ annihilation, it is even twice
inclusive: not only all produced hadronic states at a given mass are summed
over, but an integration is performed over all the possible masses from
$m_{\pi}$ to $m_{\tau}$.

This favourable situation could be spoiled by the fact that the $Q^2$ scale
is rather small, so that questions about the validity of a perturbative
approach can be raised. At least two levels are to be considered: the 
convergence of the perturbative expansion and the control of the
nonperturbative contributions. Happy circumstances make
these contributions indeed very small\cite{braaten88,narpic88}.

\subsection{Theoretical prediction for \boldmath\Rt}

The imaginary parts of the vector 
and axial-vector two-point correlation functions 
$\Pi^{(J)}_{\bar{u}d,V/A}(s)$, with the spin $J$ of the hadronic 
system, are proportional to the $\tau$ hadronic \sfs\ with 
corresponding quantum numbers. The non-strange ratio \Rt\
can be written as an integral of these \sfs\ over the 
invariant mass-squared $s$ of the f\/inal state hadrons~\cite{bnp}:
\begin{eqnarray}
\label{eq_rtauth1}
   R_\tau(s_0) &=&
     12\pi S_{\rm EW}\intl_0^{s_0}\frac{ds}{s_0}\left(1-\frac{s}{s_0}
                                    \right)^{\!\!2} \\
   & & \hspace{-1.7cm} 
     \times \left[\left(1+2\frac{s}{s_0}\right){\rm Im}\Pi^{(1)}(s+i\e)
      \,+\,{\rm Im}\Pi^{(0)}(s+i\e)\right] \nonumber
\end{eqnarray}
By Cauchy's theorem the imaginary part of $\Pi^{(J)}$ is 
proportional to the discontinuity across the positive real axis. 

The energy scale $s_0$ for $s_0= m_\tau^2$ is large enough that
contributions from nonperturbative ef\/fects be small. It is therefore 
assumed that one can use the {\it Operator Product Expansion} (OPE) 
to organize perturbative and nonperturbative contributions~\cite{svz} to 
\Rts.

The theoretical prediction of the vector and axial-vector
ratio \RtVA\ can thus be written as:
\begin{eqnarray}
\label{eq_delta}
   R_{\tau,V/A} \;=\;
     \frac{3}{2}|V_{ud}|^2S_{\rm EW} \\
   & & \hspace{-3.7cm}
     \times \left(1 + \delta^{(0)} + 
     \delta^\prime_{\rm EW} + \delta^{(2-\rm mass)}_{ud,V/A} + 
     \hm\hm\sum_{D=4,6,8}\hm\hm\hm\hm\delta_{ud,V/A}^{(D)}\right) \nonumber
\end{eqnarray}
with the residual non-logarithmic electroweak
correction $\delta^\prime_{\rm EW}=0.0010$~\cite{braaten}, 
neglected in the following, and the dimension $D=2$ 
contribution $\delta^{(2-\rm mass)}_{ud,V/A}$ 
from quark masses which is lower than $0.1\%$ for $u,d$ quarks.
The term $\delta^{(0)}$ is the purely perturbative 
contribution, while the $\delta^{(D)}$ are the OPE
terms in powers of $s_0^{-D/2}$ of the following form
\beq
\label{eq_ope}
    \delta_{ud,V/A}^{(D)} \;\sim\;
       \hm\hm\hm\sum_{{\rm dim}{\cal O}=D}\hm\hm\hm
            \frac{\langle{\cal O}_{ud}\rangle_{V/A}}
                 {(-s_0)^{D/2}}
\eeq
where the long-distance nonperturbative ef\/fects are absorbed into
the vacuum expectation elements $\langle{\cal O}_{ud}\rangle$.

The perturbative expansion (FOPT) is known to third order~\cite{3loop}.
A resummation of all known higher order 
logarithmic integrals improves the convergence 
of the perturbative series (contour-improved method \FOPTCI\}~\cite{pert}. 
As some ambiguity persists, the results are given as an average of the
two methods with the difference taken as systematic uncertainty.

\subsection{Measurements}

The ratio \Rt\ is obtained from measurements of the leptonic branching 
ratios:
\beqn
  R_\tau&=&3.647\pm0.014 
\eeqn
using a value 
$B(\tau^-\rightarrow e^-\,\bar{\nu}_e\nu_\tau)=(17.794\pm0.045)\%$
which includes the improvement in accuracy provided by the 
universality assumption of leptonic currents together with the 
measurements of $B(\tau^-\rightarrow e^-\,\bar{\nu}_e\nu_\tau)$,
$B(\tau^-\rightarrow \mu^-\,\bar{\nu}_\mu\nu_\tau)$ and the $\tau$ 
lifetime. The nonstrange part of \Rt\ is obtained subtracting
out the measured strange contribution (see last section).

Two complete analyses of the $V$ and $A$ parts have been performed by
ALEPH~\cite{aleph_a} and OPAL~\cite{opal}. Both use the world-average
leptonic branching ratios, but their own measured spectral functions.
The results on $\alpha_s(m_\tau^2)$ are therefore strongly correlated
and indeed agree when the same theoretical prescriptions are used. 
Here only the ALEPH results are given.

\subsection{Results of the f\/its}

The results of the fits are given in table~\ref{tab_asresults}.
The limited number of observables
and the strong correlations between the spectral moments introduce
large correlations, especially between the f\/itted 
nonperturbative operators.

\begin{table}
{\small
  \begin{tabular}{|l||c|c|} \hline 
     & &  \\
 ALEPH&$\alpha_s(m_{\tau}^2)$       &$\delta_{\rm NP}$\\
\hline 
 V    &  $0.330\pm0.014\pm0.018$    &  $0.020\pm0.004$  \\
 A    &  $0.339\pm0.013\pm0.018$    &  $-0.027\pm0.0004$  \\
\hline
 V+A  &  $0.334\pm0.007\pm0.021$    &  $-0.003\pm0.0004$  \\
\hline
  \end{tabular}
}
  \caption{
              F\/it results of \asm\ and the OPE nonperturbative 
              contributions from vector, axial-vector and $(V+A)$ combined
              fits using the corresponding ratios \Rt\ and the spectral 
              moments as input parameters. The second error is given for 
              theoretical uncertainty.}
\label{tab_asresults}
\end{table}

One notices a remarkable agreement within statistical errors
between the \asm\ values using vector and axial-vector data.
The total nonperturbative power contribution to \RtVpA\ is compatible 
with zero within an uncertainty of 0.4\pc, that is much smaller than 
the error arising from the perturbative term. This cancellation of the 
nonperturbative terms increases the confidence on the \asm\ 
determination from the inclusive $(V+A)$ observables.

The f\/inal result is : 
\beq
\label{eq_asres}
   \alpha_s(m_\tau^2)    = 0.334 \pm 0.007_{\rm exp} 
                                 \pm 0.021_{\rm th} 
\eeq
where the f\/irst error accounts for the experimental uncertainty and 
the second gives the uncertainty of the theoretical prediction of
$R_\tau$ and the spectral moments as well as the ambiguity of the 
theoretical approaches employed. 

\subsection{Test of the running of \boldmath$\alpha_s(s)$ at low energies}

Using the \sfs, one can simulate the physics of a hypothetical 
$\tau$ lepton with a mass $\sqrt{s_0}$ smaller than $m_\tau$
through equation~(\ref{eq_rtauth1}) and hence further investigate QCD
phenomena at low energies. Assuming quark-hadron duality, 
the evolution of $R_\tau(s_0)$ provides a direct test of the 
running of $\alpha_s(s_0)$, governed by the RGE $\beta$-function. 
On the other hand, it is a test of the validity of the OPE approach 
in $\tau$ decays. 

\smallskip
\FIGURE{\epsfig{file=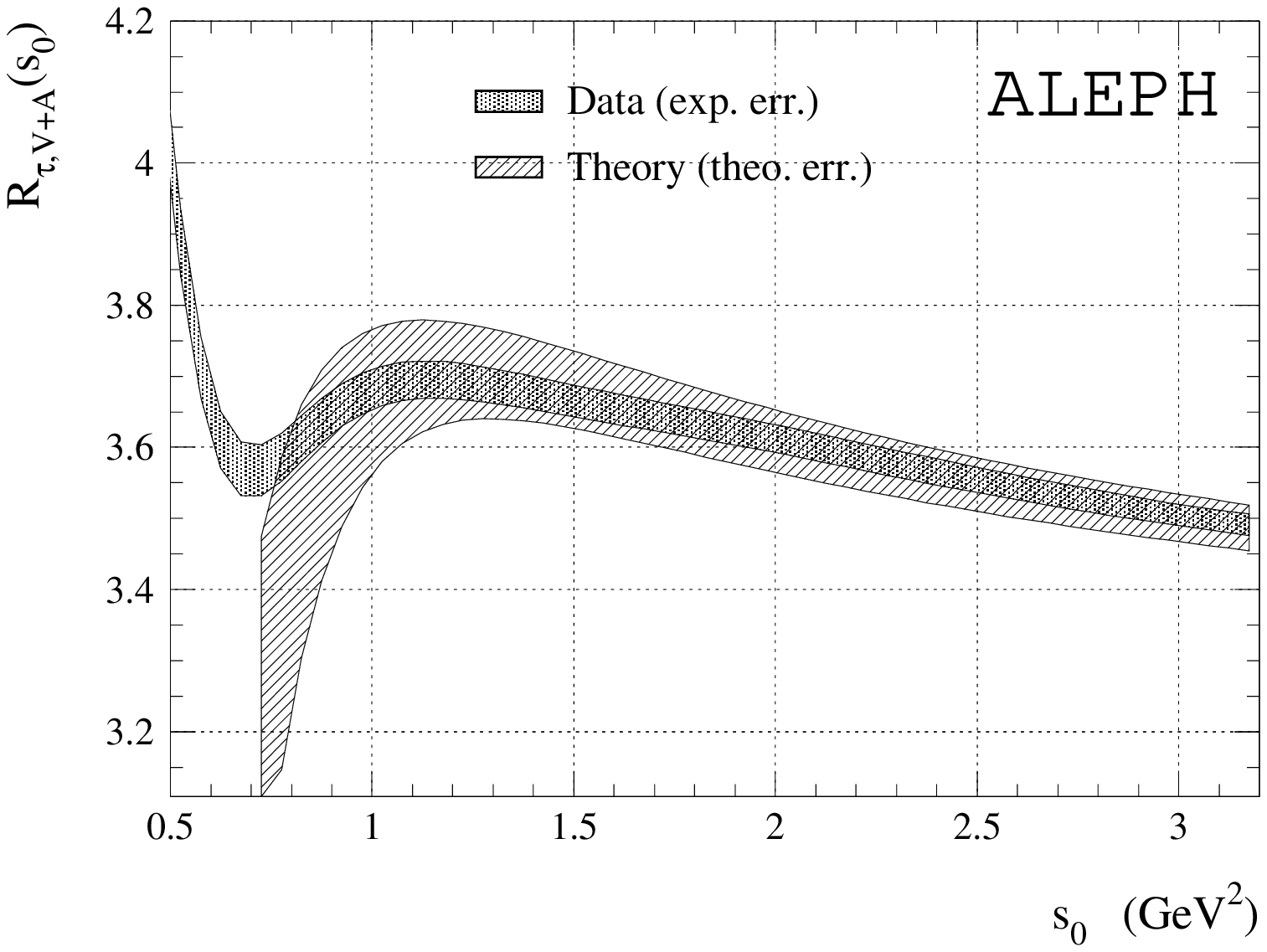,width=6cm}%
        \caption{
              The ratio \RtVpA\ versus the square ``$\tau$ mass'' $s_0$. 
              The curves are plotted as error bands to emphasize their 
              strong point-to-point correlations in $s_0$. Also 
              shown is the theoretical prediction using 
              the results of the fit at $s_0=m_\tau^2$.}%
	\label{fig_rtau}}

\smallskip
\FIGURE{\epsfig{file=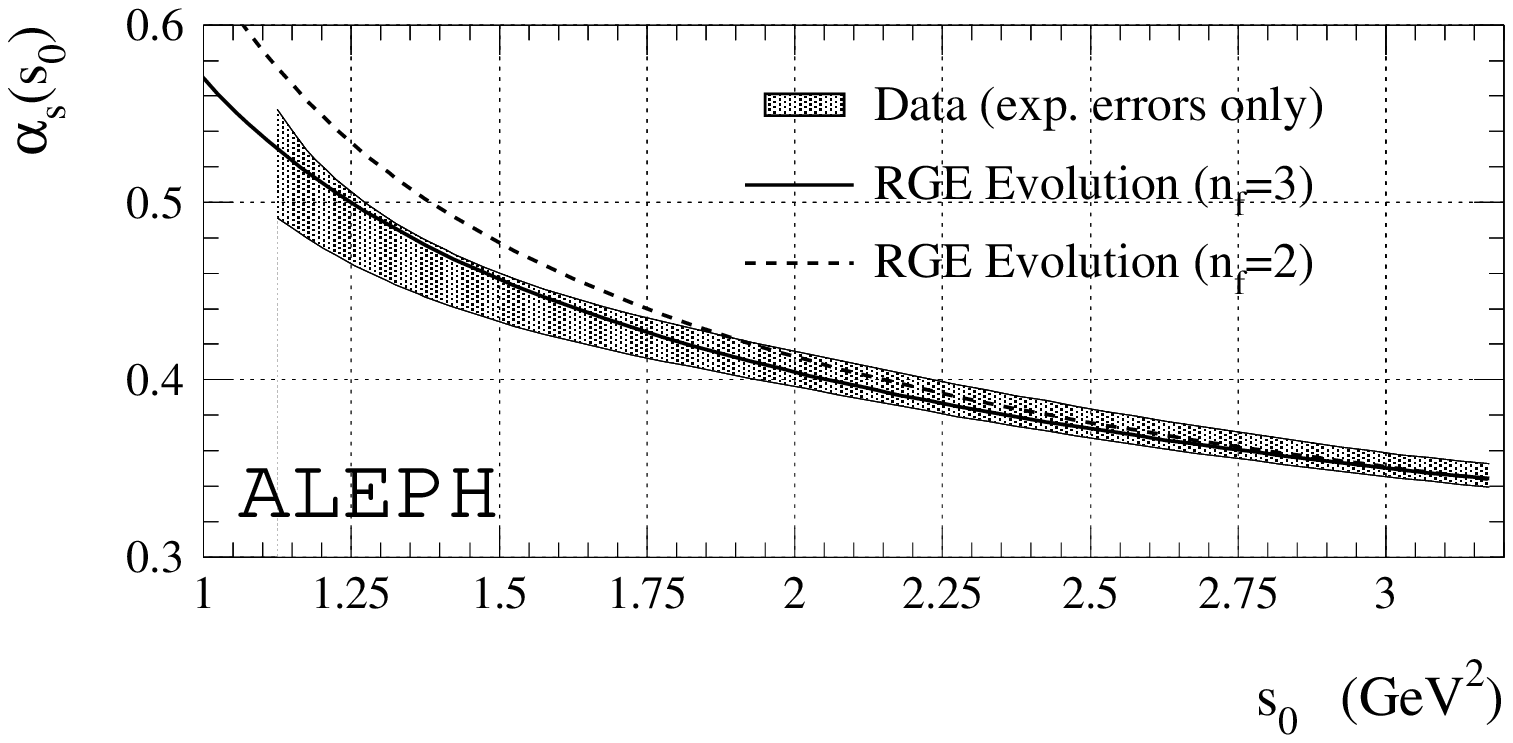,width=6cm}%
        \caption{
              The running of $\alpha_s(s_0)$ obtained from the 
              fit of the theoretical prediction to \RtVpAs.
              The shaded band shows the data including experimental
              errors. The curves give the four-loop RGE evolution 
              for two and three flavours.}%
	\label{fig_runas}}

The functional dependence of \RtVpAs\ is plotted in 
figure~\ref{fig_rtau} together with the theoretical 
prediction using the results of table~\ref{tab_asresults}.
Below $1~{\rm GeV}^2$ the error of the theoretical 
prediction of \RtVpAs\ starts to blow up due to the 
increasing uncertainty from the unknown fourth-order perturbative term.
F\/igure~\ref{fig_runas} shows the plot 
corresponding to F\/ig.~\ref{fig_rtau}, translated into the running 
of $\alpha_s(s_0)$, \ie, the experimental value for $\alpha_s(s_0)$ 
has been individually determined at every $s_0$ from the comparison 
of data and theory. Good agreement is observed with the four-loop 
RGE evolution using three quark f\/lavours.

The experimental fact that the nonperturbative contributions 
cancel over the whole range $1.2~{\rm GeV}^2\le s_0\le m_\tau^2$ 
leads to conf\/idence that the \as\ determination from the inclusive 
$(V+A)$ data is robust.  

\subsection{Discussion on the determination of $\alpha_s(m_\tau^2)$}

The evolution of the 
\asm\ measurement from the inclusive $(V+A)$ observables based on 
the Runge-Kutta integration of the dif\/ferential equation 
of the renormalization group to 
N$^3$LO~\cite{alpha_evol,pichsanta} yields
\beq
\label{alphaevol}  
   \alpha_s(M_{\rm Z}^2) = \nonumber \\
                           0.1202 \pm 0.0008_{\rm exp} 
                                  \pm 0.0024_{\rm th} 
                                  \pm 0.0010_{\rm evol}
\eeq
where the last error stands for
possible ambiguities in the evolution due to uncertainties in the 
matching scales of the quark thresholds~\cite{pichsanta}. 

\smallskip
\FIGURE{\epsfig{file=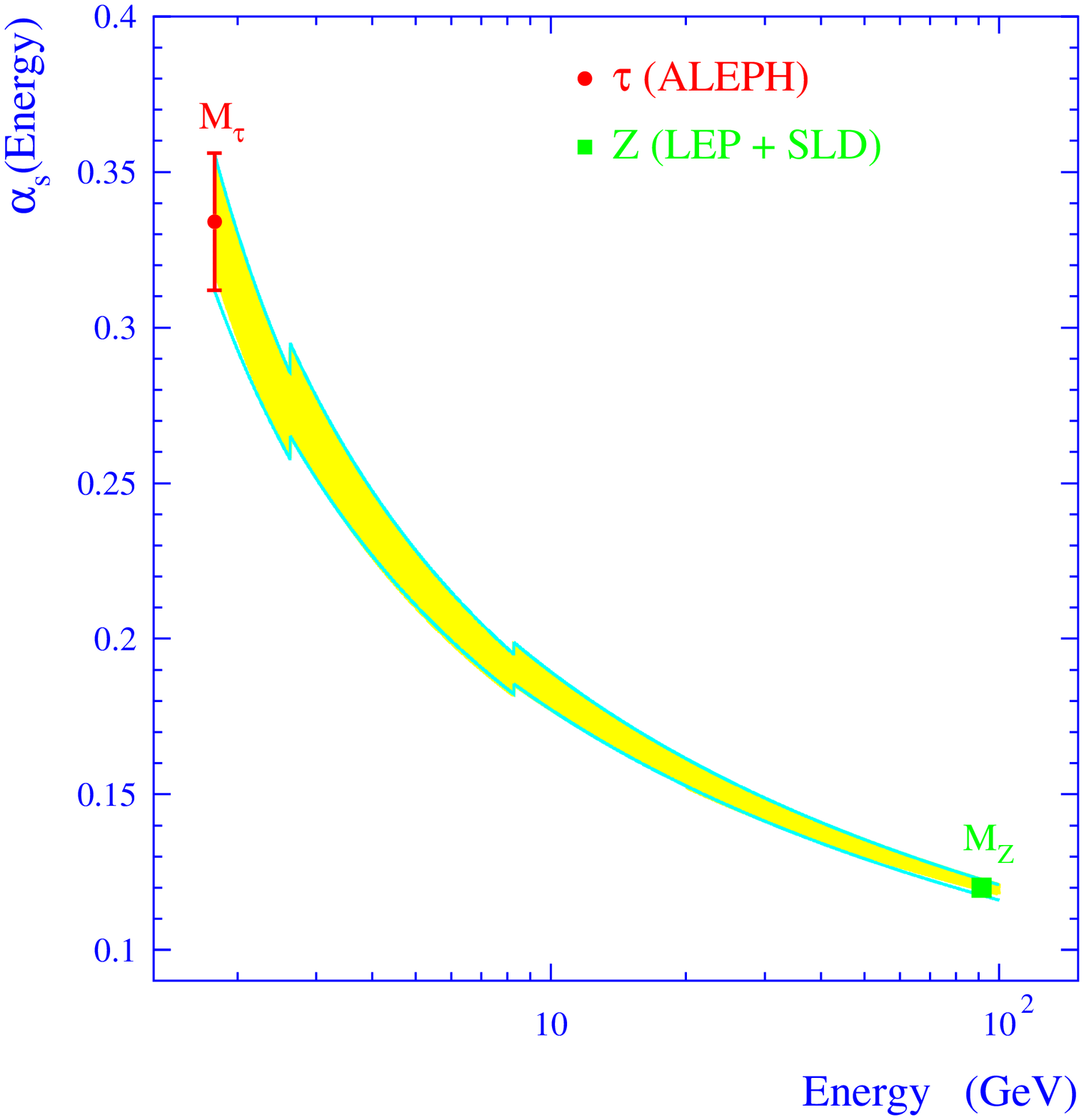,width=5cm}%
        \caption{
     Evolution of the strong coupling measured at $m_\tau^2)$
     to $M_Z^2$ predicted by QCD compared to the direct measurement.
     The evolution is carried out at 4 loops, while the flavour
     matching is accomplished at 3 loops at $2~m_c$ and $2~m_b$
     thresholds.}%
	\label{alphatz}}

The result (\ref{alphaevol}) can be compared to the
determination from the global electroweak fit. 
The variable $R_Z$ has similar advantages as $R_\tau$, but 
it differs concerning the convergence of the perturbative
expansion because of the much larger scale. It turns out that this
determination is dominated by experimental errors with very small
theoretical uncertainties, \ie\ the reverse of the situation 
encountered in $\tau$ decays.
The most recent value\cite{blondel_blois} yields 
$\alpha_s(M_{\rm Z}^2) = 0.119 \pm 0.003$, in excellent agreement
with (\ref{alphaevol}). 
Fig.~\ref{alphatz} illustrates well the agreement between the evolution
of $\alpha_s(m_{\tau}^2)$ predicted by QCD and
$\alpha_s(M_{\rm Z}^2)$.

\section{Applications to hadronic vacuum polarization}

\subsection{Improvements to the standard calculations}

From the studies presented above we have learned that:
\begin{itemize}
\item the $I=1$ vector \sf\ from $\tau$ decays agrees with that from
\ee\ annihilation, while it is more precise for masses less than
$1.6$~GeV as can be seen on figure~\ref{cvc_aleph}.
Small CVC violations are expected at a few $10^{-3}$ level
\cite{alemanyhd} from radiative $\rho$ decays and SU(2)-breaking in
the $\pi$ and $\rho$ masses.
\item the description of $R_\tau$ by perturbative QCD works down to 
a scale of $1$~GeV. Nonperturbative contributions at $1.8~GeV$ are well
below 1 \% in this case. They are larger ($\sim 3$ \%) 
for the vector part alone, but reasonably well 
described by OPE. The complete (perturbative + nonperturbative)
description is accurate at the 1 \% level at $1.8$~GeV for
integrals over the vector \sf\, such as $R_{\tau,V}$.
\end{itemize}

\smallskip
\FIGURE{\epsfig{file=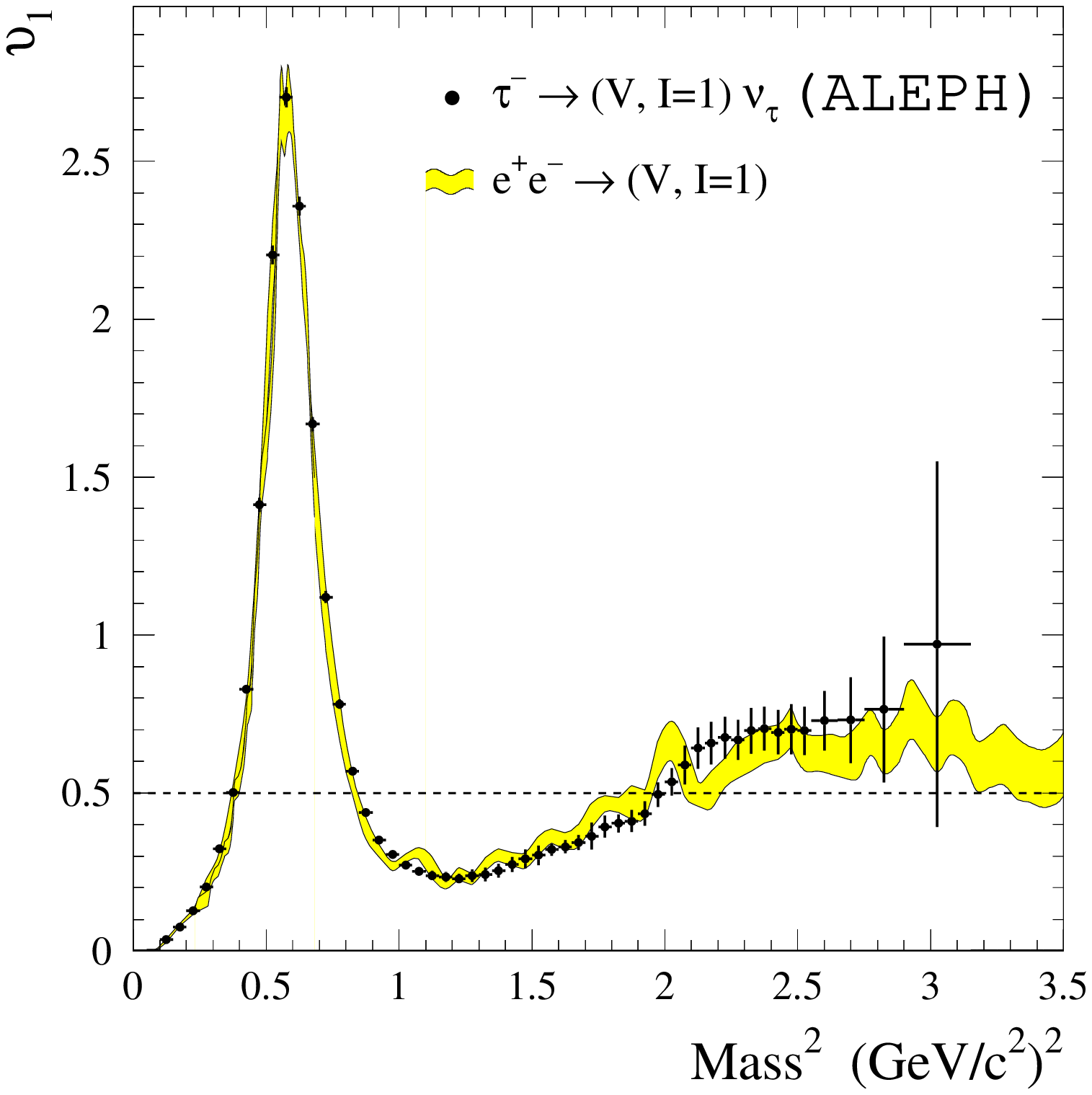,width=6cm}%
        \caption{Global test of CVC using $\tau$ and $e^+e^-$ vector
         spectral functions.}%
	\label{cvc_aleph}}

These two facts have direct applications to calculations of
hadronic vacuum polarisation which involve the knowledge of the vector
\sf: the muon magnetic anomaly and the running of $\alpha$. In both cases, 
the standard method involves a dispersion integral over
the vector \sf\ taken from the $e^+e^-~\rightarrow$ hadrons
data. Eventually at large energies, QCD is used to replace
experimental data. Hence the precision of the calculation is given
by the accuracy of the data, which is poor above $1.5$~GeV. 
Even at low energies, the precision
can be significantly improved at low masses by using $\tau$ data
\cite{alemanyhd}.

The next breakthrough comes about using the prediction of perturbative QCD 
far above quark thresholds, but at low enough energies (compatible with the
remarks above) in place of noncompetitive experimental data\cite{hd1}. This 
procedure involves a proper treatment of the quark masses in the QCD
prediction\cite{chet1}.

Finally, it is still possible to improve the contributions from data by
using analyticity and QCD sum rules, basically without any additional
assumption. This idea, advocated in Ref.~\cite{schilcher}, has been used
within the procedure described above to still improve the calculations
\cite{hd2}.

The experimental results of $R(s)$ 
and the theoretical prediction are shown in figure~\ref{rdata}. 
The shaded bands depict the regions where data are used instead 
of theory to evaluate the respective integrals. Good agreement 
between data and QCD is found above 8~GeV, while at lower energies 
systematic deviations are observed. The $R$ measurements in this 
region are essentially provided by the $\gamma\gamma2$~\cite{E_78} 
and MARK~I~\cite{E_96} collaborations. MARK~I data above 5~GeV lie 
systematically above the measurements of the Crystal Ball~\cite{CB} 
and MD1~\cite{MD1} Collaborations as well as the QCD prediction.

\smallskip
\FIGURE{\epsfig{file=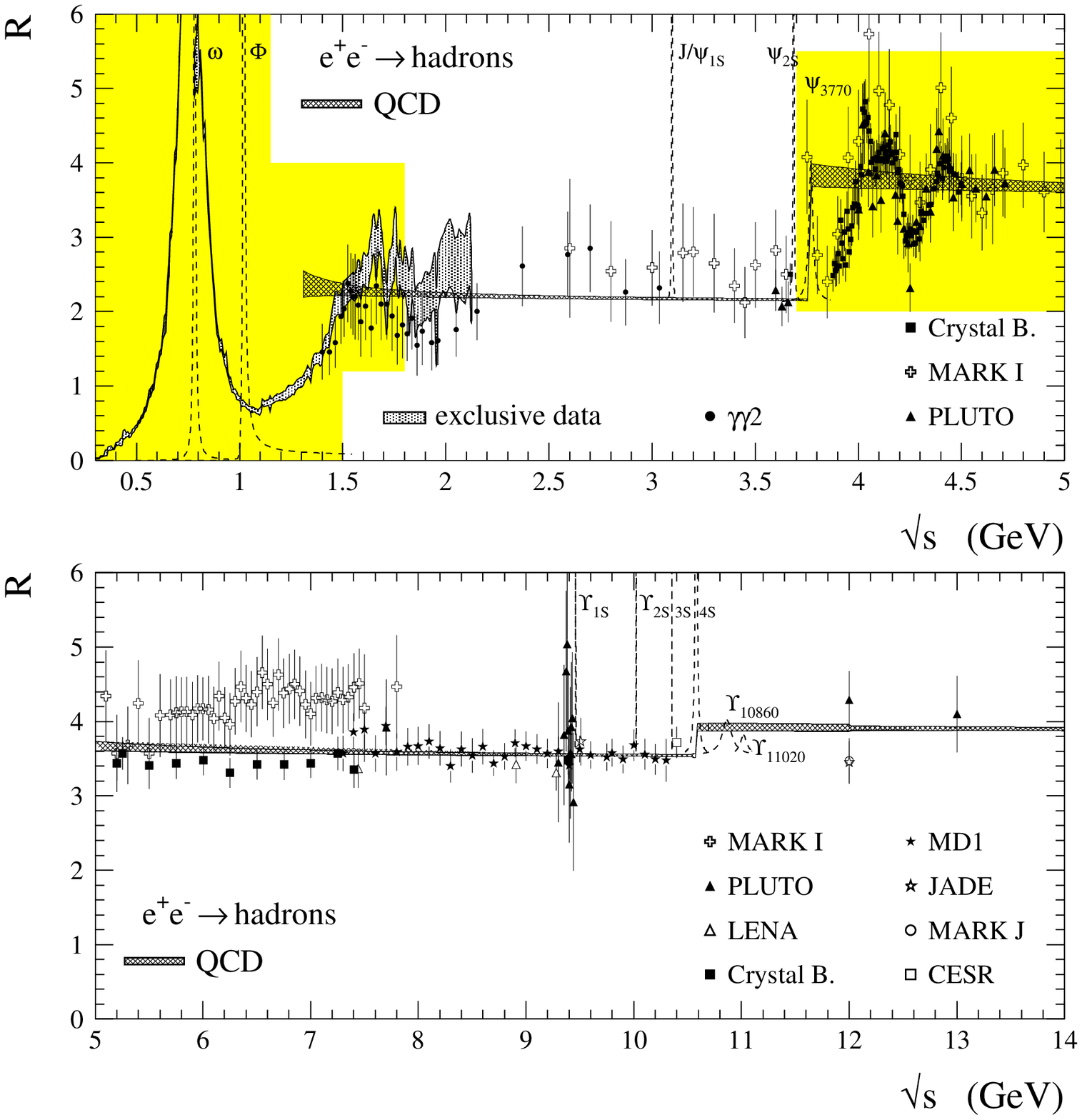,width=7cm}%
        \caption{
            Inclusive hadronic cross section ratio in \ee\
            annihilation versus the c.m. energy $\sqrt{s}$. 
            Additionally shown is the QCD prediction of the continuum 
            contribution from Ref. \cite{hd1}
            as explained in the text. The shaded areas 
            depict regions were experimental data are used for the 
            evaluation of \daqedhZ\ and \amuhad\ in addition to the 
            measured narrow resonance parameters. The exclusive 
            \ee\ cross section measurements at low c.m. energies
            are taken from DM1,DM2,M2N,M3N,OLYA,CMD,ND and
            $\tau$ data from ALEPH (see Ref.~\cite{alemanyhd}
            for detailed information).}%
	\label{rdata}}

\subsection{Muon magnetic anomaly}

By virtue of the analyticity of the vacuum polarization correlator, the 
contribution of the hadronic vacuum polarization to $a_\mu$ can be 
calculated \via\ the dispersion integral~\cite{rafael}
\beq\label{eq_integral1}
    a_\mu^{\mathrm had} \:=\: 
           \frac{1}{4\pi^3}
           \intl_{4m_\pi^2}^\infty ds\,\sigma_{\mathrm had}(s)\,K(s)
\eeq
Here $\sigma_{\mathrm had}(s)$ is the total \ee$\rightarrow\,$hadrons 
cross section as a function of the c.m. energy-squared $s$, and
$K(s)$ denotes a well-known QED kernel.

The function $K(s)$ decreases monotonically with increasing $s$. It gives
a strong weight to the low energy part of the integral~(\ref{eq_integral1}).
About 91\pc\ of the total contribution to \amuhad\ is accumulated at c.m. 
energies $\sqrt{s}$ below 2.1~GeV while 72\pc\ of \amuhad\ is covered by 
the two-pion f\/inal state which is dominated by the $\rho(770)$ resonance. 
The new information provided by the ALEPH 2- and 4-pion \sfs\ can
signif\/icantly improve the \amuhad\ determination.

\subsection{Running of the electromagnetic coupling}

In the same spirit we evaluate the hadronic contribution \daqed\
to the renormalized vacuum polarization function $\Pi_\gamma^\prime(s)$ 
which governs the running of the electromagnetic coupling 
\aqed.  With \daqed=$-4\pi\alpha\,{\mathrm Re}
\left[\Pi_\gamma^\prime(s)-\Pi_\gamma^\prime(0)\right]$, one has
\beq
    \alpha(s) \:=\: \frac{\alpha(0)}{1-\Delta\alpha(s)}
\eeq
where $4\pi\alpha(0)$ is the square of the electron charge in the 
long-wavelength Thomson limit.

The leading order leptonic contribution is equal to $314.2\times10^{-4}$.
Using analyticity and unitarity, the dispersion integral for the contribution
from the light quark hadronic vacuum polarization \daqedhZ\ reads~\cite{cabibbo}
\beq\label{eq_integral2}
    \Delta\alpha_{\mathrm had}^{(5)}(M_{\mathrm Z}^2) \:=\: \nonumber \\
        -\frac{M_{\mathrm Z}^2}{4\pi^2\,\alpha}\,
         {\mathrm Re}\intl_{4m_\pi^2}^{\infty}ds\,
            \frac{\sigma_{\mathrm had}(s)}
                 {s-M_{\mathrm Z}^2-i\epsilon}
\eeq
where $\sigma(s)=16\pi^2\alpha^2(s)/s\cdot{\mathrm Im}\Pi_\gamma^\prime(s)$
from the optical theorem. In contrast to 
\amuhad, the integration kernel favours cross sections at higher masses. 
Hence, the improvement when including $\tau$ data is expected to be small.

The top quark contribution can be calculated using the next-to-next-to-leading
order $\alpha_s^3$ prediction of the total inclusive cross section ratio $R$
from perturbative QCD~\cite{3loop,eidelman}. The evaluation of the 
integral~(\ref{eq_integral2}) with $m_{\mathrm top}=175$~GeV yields 
$\Delta\alpha_{\mathrm top}(M_{\mathrm Z}^2)=-0.6\times10^{-4}$.

\subsection{Results}

The combination of the theoretical
and experimental evaluations of the integrals~(\ref{eq_integral2}) 
and (\ref{eq_integral1}) yields the results
\beqn
\label{eq_amalp}
   \Delta\alpha_{\rm had}(M_{\rm Z}^2) 
      &=& (276.3 \pm 1.1_{\rm exp} \pm 1.1_{\rm th})\times10^{-4} \nonumber\\
   \alpha^{-1}(M_{\rm Z}^2) 
      &=& 128.933 \pm 0.015_{\rm exp} \pm 0.015_{\rm th} \nonumber\\ 
   a_\mu^{\rm had}
      &=& (692.4 \pm 5.6_{\rm exp} \pm 2.6_{\rm th})\times10^{-10} \nonumber\\
   a_\mu^{\rm SM}
      &=& (11\,659\,159.6 \nonumber\\
   & & \hspace{0cm} 
            \pm 5.6_{\rm exp} \pm 3.7_{\rm th})\times10^{-10}
\eeqn
and $a_e^{\rm had}=(187.5\pm1.7_{\rm exp}\pm0.7_{\rm th})\times10^{-14}$ 
for the leading order hadronic contribution to $a_e$.
The total $a_\mu^{\rm SM}$ value includes additional contributions
from non-leading order hadronic vacuum polarization (summarized in
Refs.\cite{krause2,alemanyhd}) and light-by-light scattering~\cite
{kinolight,light2} contributions.  
Figures~\ref{fig_results_alpha} and \ref{fig_results_amu} show a 
compilation of published results for the hadronic contributions 
to \aqedZ\ and $a_\mu$. Some authors give the hadronic contribution 
for the f\/ive light quarks only and add the top quark part separately. 
This has been corrected for in F\/ig.~\ref{fig_results_alpha}.

\smallskip
\FIGURE{\epsfig{file=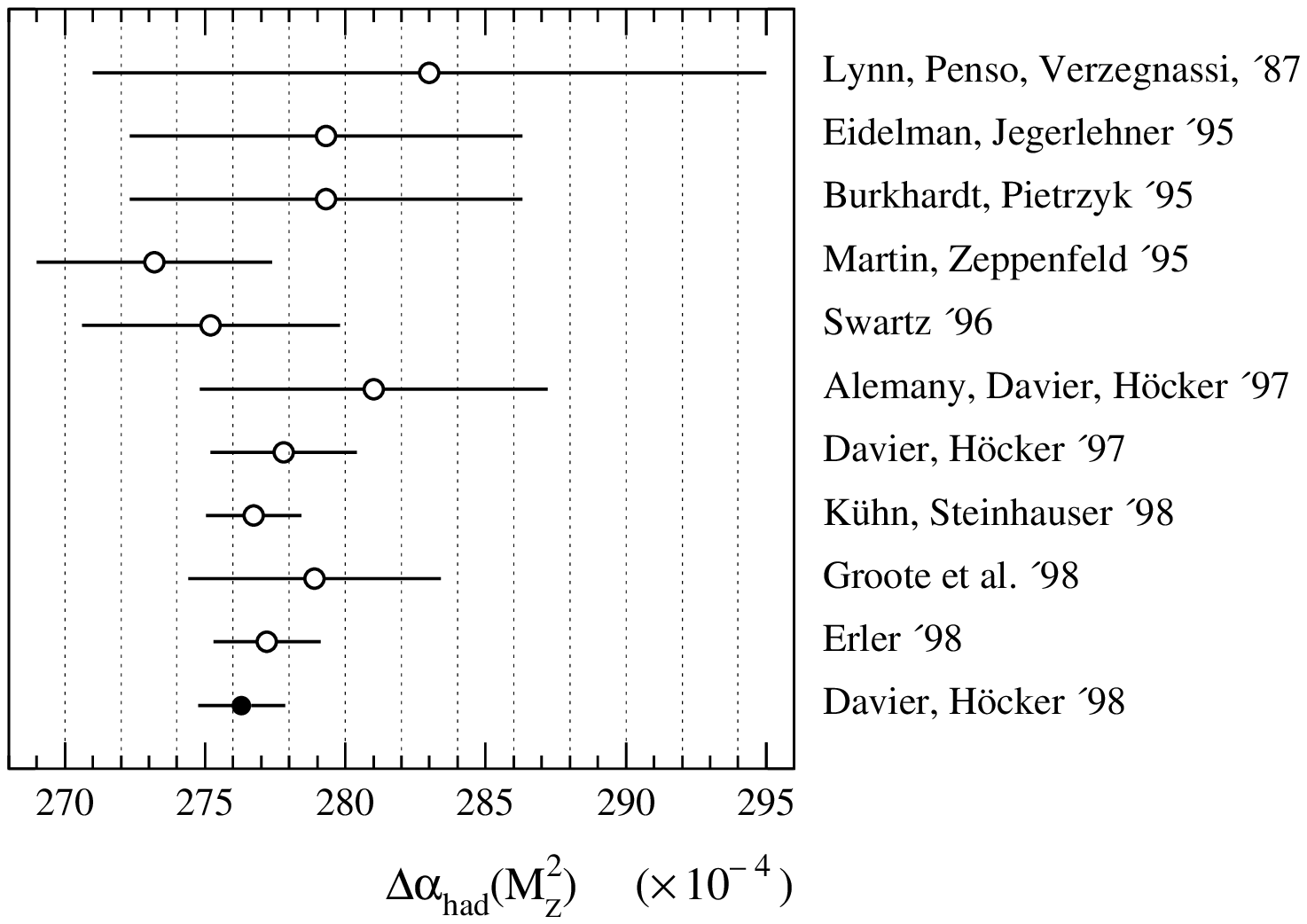,width=7cm}%
        \caption{
            Comparison of \daqedhZ\ evaluations. The values are
            taken from Refs.~\rm\cite{lynn,eidelman,burkhardt,martin,swartz,
            alemanyhd,hd1,kuhnstein,erler,hd2}.}%
	\label{fig_results_alpha}}

\smallskip
\FIGURE{\epsfig{file=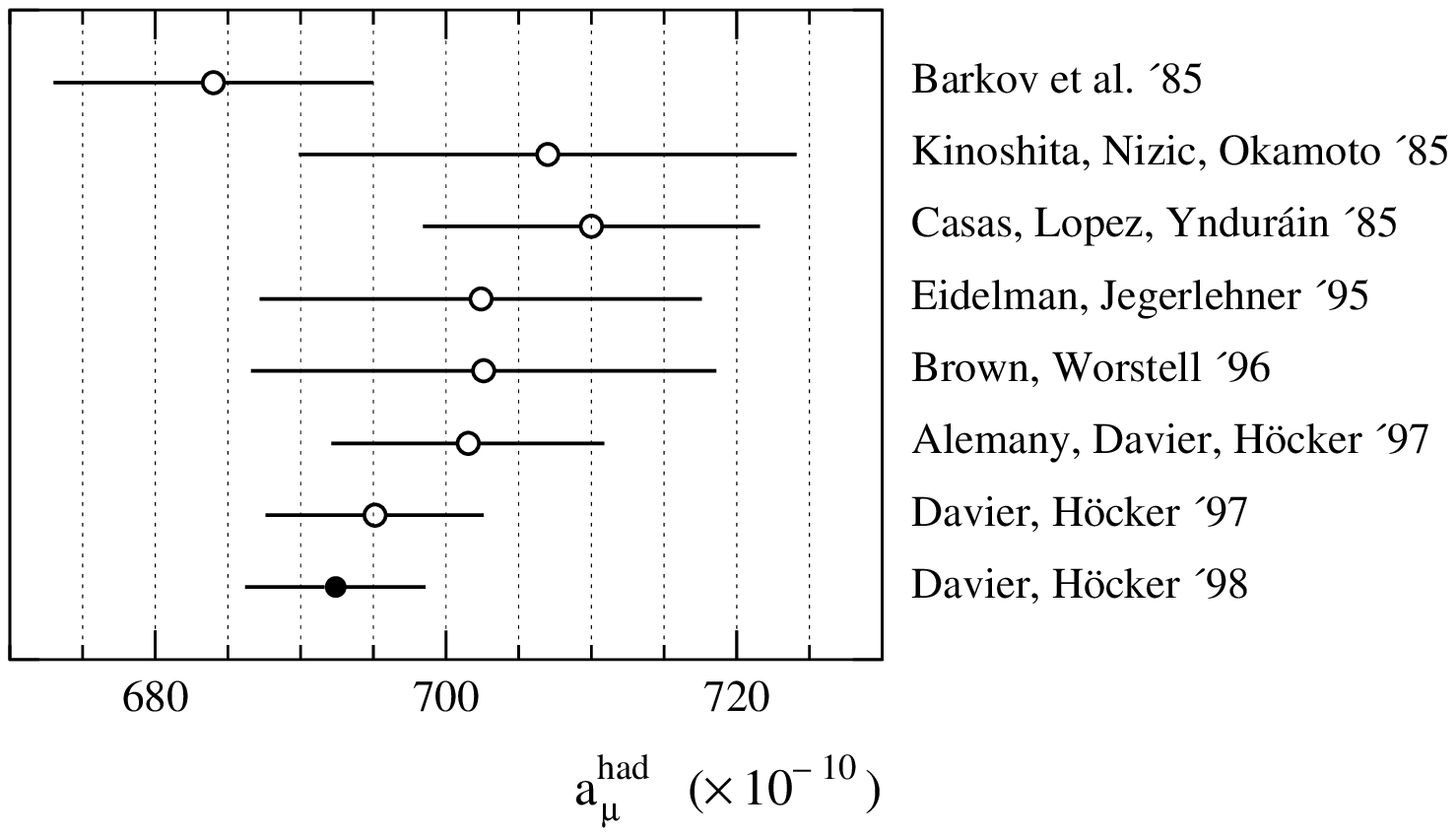,width=7cm}%
        \caption{
            Comparison of $a_\mu^{\rm had}$ evaluations. The values are
            taken from Refs.~\rm\cite{barkov,kinoshita,casas,eidelman,
            worstell,alemanyhd,hd1,hd2}.}%
	\label{fig_results_amu}}

\subsection{Outlook}

These results have direct implications for phenomenology and
on-going experimental programs.
 
Most of the sensitivity to the Higgs boson mass 
originates from the measurements of asymmetries in the process 
$e^+e^- \rightarrow Z \rightarrow$ fermion pairs 
and {\it in fine} from $(sin^2 \theta_W)_{eff} = \bar s^2$.
Unfortunately, this approach is limited by the fact that the intrinsic
uncertainty on $\alpha (M_{\rm Z}^2)$ in the standard evaluation
is at the same level as the experimental accuracy on $\bar s^2$.
The situation has completely changed with the new determination of \aqedZ\ 
which does not limit anymore the adjustment of the Higgs mass from 
accurate experimental determinations of 
${\rm sin}^2\theta_{\rm W}$.
The improvement in precision can be directly appreciated on the
relevant variable $log~M_H$ with $M_H$ in ${\rm GeV}/c^2$~\cite{blondel_blois}:
\begin{eqnarray}
log~M_H &=& 1.88^{+0.31}_{-0.39}
\end{eqnarray}
with the 'standard' $\alpha(M_Z^2)$~\cite{eidelman}, and
\begin{eqnarray}
log~M_H &=& 1.97^{+0.22}_{-0.25}
\end{eqnarray}
with the QCD-improved value~\cite{hd2}.

The interest in reducing the uncertainty in the hadronic contribution
to $a_\mu^{\rm had}$ is directly linked to the possibility of measuring the
weak contribution:
\beq
    a_\mu^{\mathrm SM} \:=\: a_\mu^{\mathrm QED} + a_\mu^{\mathrm had} +
                             a_\mu^{\mathrm weak}
\eeq
where $a_\mu^{\mathrm QED}=(11\,658\,470.6\,\pm\,0.2)\times10^{-10}$ is 
the pure electromagnetic contribution (see~\cite{krause1} and references 
therein), \amuhad\ is the contribution from hadronic vacuum polarization,
and $a_\mu^{\mathrm weak}=(15.1\,\pm\,0.4)\times10^{-10}
$~\cite{krause1,kuraev,weinberg} accounts for corrections due to the 
exchange of the weak interacting bosons up to two loops.
The present value from the combined 
$\mu^+$ and $\mu^-$ measurements~\cite{bailey},
\beq
    a_\mu \:=\: (11\,659\,230 \pm 85)\times10^{-10}
\eeq
should be improved to a precision of at least
$4\times10^{-10}$ by a forthcoming Brookhaven experiment (BNL-E821)~\cite{bnl},
well below the expected weak contribution. Such a programme makes sense only if
the uncertainty on the hadronic term is made sufficiently small. The 
improvements described above represent a significant step in this direction.

\section{Strange $\tau$ decays and $m_s$}

\subsection{The strange hadronic decay ratio \RtS}

As previously demonstrated in Ref.~\cite{davchen}, the inclusive 
$\tau$ decay ratio into strange hadronic final states,
\beq
     R_{\tau,S} =
\frac{\Gamma(\tau^-\rightarrow{\rm hadrons}_{S=-1}^-\,\nu_\tau)}
                   {\Gamma(\tau^-\rightarrow e^-\,\bar{\nu}_e\nu_\tau)}
\eeq
can be used due to its precise theoretical prediction~\cite{bnp,chetkwiat} 
to determine \mss\ at the scale $s=M_\tau^2$. 
Since then it was shown~\cite{maltman} that 
the perturbative expansion used for the massive 
term in Ref.~\cite{chetkwiat} was incorrect. After correction the series
shows a problematic convergence behaviour~\cite{maltman,pichprad,kuhnchet}.

\smallskip
\FIGURE{\epsfig{file=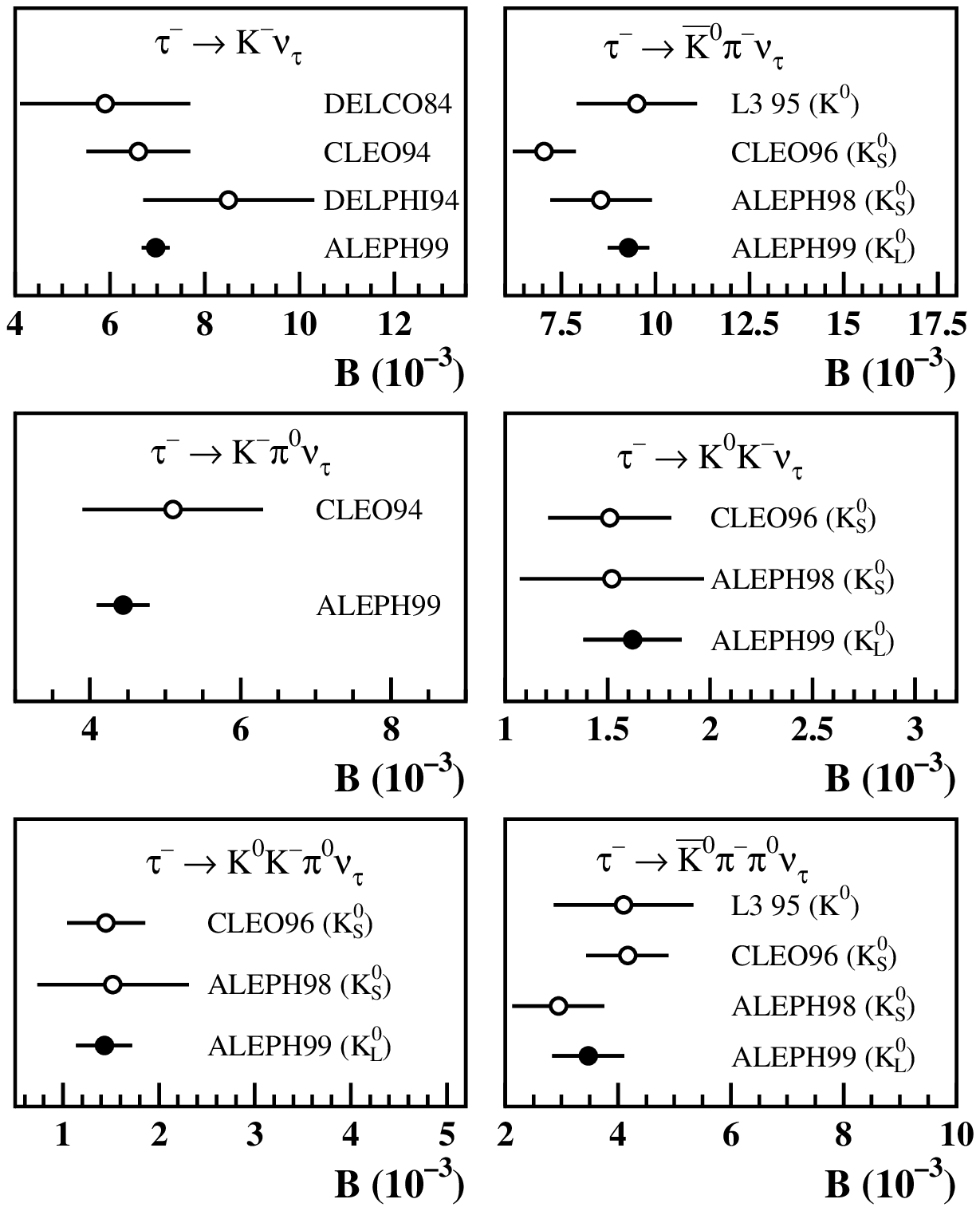,width=7cm}%
        \caption{Some of the new ALEPH results on decay modes with kaons
         compared to the published data.}%
	\label{aleph_k}}

Similarly to the nonstrange case, the QCD prediction is given by 
equations (\ref{eq_rtauth1}) and (\ref{eq_delta}) where the attention is now 
turned to the $\delta^{(2-\rm mass)}$ term, important for the relatively 
heavy strange quark. The corresponding perturbative expansion is known to
second order for the $J=1+0$ part and to the third order for $J=0$~\cite
{chetkwiat,maltman}. While the $J=1+0$ series behaves well, the $J=0$ 
expansion in fact diverges after the second term.

Following these observations, two methods can be considered 
in order to determine $m_s(m_\tau^2)$:
\begin{itemize}
\item in the {\it inclusive method}, the inclusive strange
hadronic rate is considered and both $J=1+0$ and $J=0$ are included
with their respective convergence behaviour taken into account
in the theoretical uncertainties.

\item the {\it `1+0' method} singles out the 
well-behaved $J=1+0$ part by subtracting the experimentally
determined $J=0$ longitudinal component from data. The measurement is then
less inclusive and the sensitivity to $m_s$ is significantly
reduced; however, the $\delta_2$ perturbative expansion is under
control and the corresponding theoretical uncertainty is reduced.
\end{itemize}

\subsubsection{New ALEPH results on strange decays}

ALEPH has recently published a comprehensive study of $\tau$ decay modes
including kaons (charged, $K^0_S$ and $K^0_L$)~\cite
{aleph_k3,aleph_ks,aleph_k1,aleph_k} up to four hadrons in the final state. 
A comparison with the published results is given in figure~\ref{aleph_k}.

The total \br\ for $\tau$ into strange final states, $B_S$, is
\beqn
      B_S=(2.87 \pm 0.12)\%
\eeqn
corresponding to
\beqn
      R_{\tau,S}= 0.1610 \pm 0.0066
\label{rtaus_aleph}
\eeqn
Since the QCD expectation for a massless quarks is $0.1809 \pm 0.0036$,
the result~(\ref{rtaus_aleph}) is evidence for a massive $s$ quark.

The strange spectral function is given in figure~\ref{aleph_ssf},
dominated at low mass by the $K^*$ resonance.

\smallskip
\FIGURE{\epsfig{file=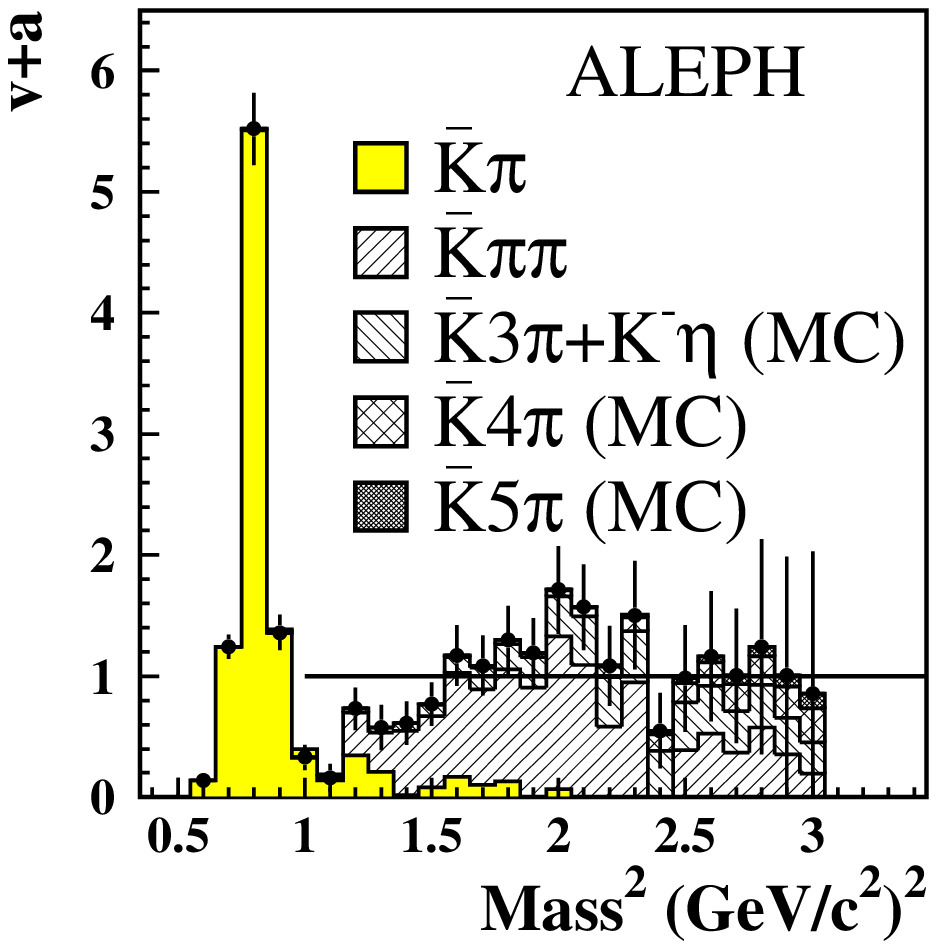,width=6cm}%
        \caption{The strange spectral function measured by ALEPH.}%
	\label{aleph_ssf}}

\subsubsection{The ALEPH analysis for \ms}
As proposed in Ref.~\cite{pichledib} and successfully applied in
\asm\ analyses, the \sf\ is used to construct moments
\beq
\label{eq_moments}
   R_{\tau,S}^{kl} \;\equiv\; 
       \intl_0^{M_\tau^2} ds\,\left(1-\frac{s}{M_\tau^2}\right)^{\!\!k}
                              \left(\frac{s}{M_\tau^2}\right)^{\!\!l}
       \frac{dR_{\tau,S}}{ds}
\eeq

In order to reduce the theoretical uncertainties one
considers the difference between non-strange and strange spectral 
moments, properly normalized with their respective CKM matrix elements:
\beq
\label{eq_dmoments}
   \Delta_\tau^{kl} \;\equiv\;
     \frac{1}{|V_{ud}|^2}R_{\tau,S=0}^{kl} - 
     \frac{1}{|V_{us}|^2}R_{\tau,S=-1}^{kl}
\eeq
for which the massless perturbative contribution vanishes so that the
theoretical prediction now reads (setting $m_u=m_d=0$)
\beq
\label{eq_Deltamom}
   \Delta_\tau^{kl} \;=\;
     3S_{\rm EW}\left(
     -\,\delta_{S}^{kl(2-\rm mass)} + 
     \hm\hm\sum_{D=4,6,\dots}\hm\hm\hm\hm\tilde{\delta}^{kl(D)}\right)
\eeq
For the
CKM matrix elements the values $|V_{ud}|=0.9751\pm0.0004$ and 
$|V_{us}|=0.2218\pm0.0016$~\cite{pdg98} are used, while the errors are
included in the theoretical uncertainties.

Figure~\ref{diffmoms} shows the weighted integrand of the lowest 
moment $\Delta^{00}_\tau$ from the ALEPH data, as 
a function of the invariant mass-squared, and for which 
the expectation from perturbative QCD vanishes.

\smallskip
\FIGURE{\epsfig{file=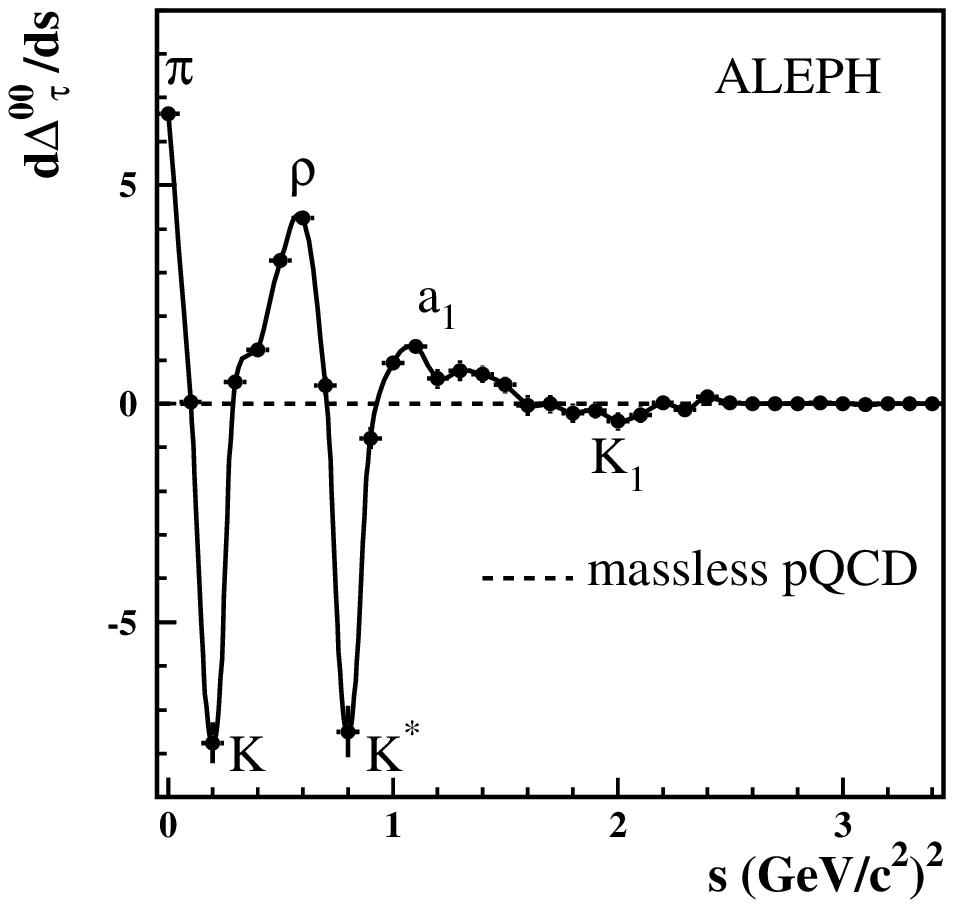,width=6cm}%
        \caption{
         Integrand of Eq.~(\ref{eq_dmoments}) for (k=0, l=0),
         $i.e.$, difference of the Cabibbo corrected non-strange and strange 
         invariant mass spectra.}%
	\label{diffmoms}}

Subtracting out the experimental $J=0$ contribution (essentially given by 
the single K channel), a fit to 5 moments is performed with the safer
$J=1+0$ method, yielding
\begin{eqnarray}
\label{res_ms}
   m_s(m_\tau^2) &=& (176^{ \,+37_{\rm exp} + 24_{\rm th}}
                   _{\,-48_{\rm exp} - 28_{\rm th}})~{\rm MeV}/c^2 \\
   \tilde{\delta}^{(6)} &=& 0.039  \pm 0.016_{\rm exp}
                           \pm 0.014_{\rm th}\\
   \tilde{\delta}^{(8)} &=& -0.021 \pm 0.014_{\rm exp}
                           \pm 0.008_{\rm th}
\end{eqnarray}
with additional (smaller) uncertainties from the fitting procedure and
the determination of the $J=0$ part. The quoted theoretical errors are
mostly from the $V_{us}$ uncertainty. The $D=6$ and $D=8$ strange 
contributions are found to be surprizingly larger than their nonstrange
counterparts. If the fully inclusive method is used instead (with the
problematic convergence) the result
$m_s(m_\tau^2)=(149^{ \,+24_{\rm exp} + 21_{\rm th}}
                   _{\,-30_{\rm exp} - 25_{\rm th}})~{\rm MeV}/c^2$
is obtained.

The result (\ref{res_ms}) can be evolved to the scale of 1 GeV 
using the four-loop RGE $\gamma$-function~\cite{ritmass}, yielding
\beq
   m_s(1~{\rm GeV}^2) = (234^{\,+61}_{\,-76})~{\rm MeV}/c^2
\eeq

This value of \ms\ is somewhat larger than previous determinations~\cite
{ms_all}, but consistent with them
within errors.
  
\section{Conclusions}
The decays $\tau \rightarrow \nu_\tau$ + hadrons constitute a clean and
powerful way to study hadronic physics up to $\sqrt{s} \sim 1.8$ GeV.
Beautiful resonance analyses have already been done, providing new
insight into hadron dynamics. Probably the major surprize has been the 
fact that inclusive hadron production is well described by perturbative
QCD with very small nonperturbative components at the $\tau$ mass.
Despite the fact that this low-energy region is dominated by resonance
physics, methods based on global quark-hadron duality work indeed very
well.

The measurement of the vector and axial-vector spectral functions has
provided the way for quantitative analyses. Precise determinations of
$\alpha_s$ agree for both spectral functions and they also agree with all
the other determinations from the Z width, the rate of Z to jets and
deep inelastic lepton scattering. Indeed from $\tau$ decays
\beqn
      \alpha_s(M_{\rm Z}^2)_\tau=0.1202 \pm 0.0027
\eeqn
in excellent agreement with the average from all other 
determinations~\cite{alphamor}
\beqn
      \alpha_s(M_{\rm Z}^2)_{non-\tau}=0.1187 \pm 0.0020
\eeqn

The use of the vector $\tau$ spectral function and the QCD-based approach
as tested in $\tau$ decays improve the calculations of hadronic vacuum
polarization considerably. Significant results have been obtained for
the running of $\alpha$ to the Z mass and the muon anomalous magnetic
moment. Both of these quantities must be known with high precision as
they give access to new physics. 

Finally the strange spectral function has been measured, providing
a determination of the strange quark mass.

\acknowledgments

The author would like to thank my ALEPH colleagues
S.M.~Chen, A.~H\"ocker and C.Z.~Yuan 
for their precious collaboration, A.~Weinstein for fruitful discussions
on the CLEO data and the organizers of Heavy Flavours 8 for a nice
conference.

\end{document}